\documentclass{article}[a4paper]
\usepackage{array, tabularx, caption, boldline}
\usepackage{graphicx}
\usepackage{makecell}
\usepackage{cellspace}
\usepackage[utf8]{inputenc}
\usepackage{lmodern,textcomp}
\usepackage{hyperref}
\hypersetup{
    colorlinks=true,
    linkcolor=blue,
    filecolor=magenta,      
    urlcolor=blue,
    citecolor=blue
}

\usepackage{float}
\usepackage{cite}
\usepackage{ifthen, mciteplus} 
\usepackage{geometry}

\usepackage{amssymb}

\usepackage{url}
\usepackage{placeins}
\newboolean{pdflatex}
\setboolean{pdflatex}{true} 

\newboolean{articletitles}
\setboolean{articletitles}{true} 
\newboolean{uprightparticles}
\setboolean{uprightparticles}{false} 
\newboolean{inbibliography}
\setboolean{inbibliography}{true} 

\usepackage{microtype}
\usepackage{lineno}  
\usepackage{xspace} 
\usepackage{caption} 

\newcommand{\BFBcTauNu}{\mathcal{B}(B_c^+ \to \tau^+ \nu_\tau)}
\newcommand{\BFBTauNu}{\mathcal{B}(B^+ \to \tau^+ \nu_\tau)}
\newcommand{\BTauNu}{B^+ \to \tau^+ \nu_\tau}
\newcommand{\BcTauNu}{B_c^+ \to \tau^+ \nu_\tau}
\newcommand{\TauThreePi}{\tau^+ \to \pi^+ \pi^+ \pi^- \bar{\nu}_\tau}

\title{Prospects for $B_{c}^+\to \tau^+ \nu_\tau$ at FCC-ee}
\author{Yasmine Amhis$^1$, Marie Hartmann$^2$, Cl\'ement Helsens$^{3}$, Donal Hill$^{4}$, Olcyr Sumensari$^{1}$ }
\date{}

\usepackage{xspace} 
\usepackage{upgreek}
\usepackage{amsmath}

\def\delphes {\mbox{\textsc{DELPHES}}\xspace}
\def\kSimDelphes{\mbox{\textsc{k4SimDelphes}}\xspace}
\def\edmhep{\mbox{\textsc{EDM4hep}}\xspace}
\def\fasttrack{\mbox{\textsc{FastTrackCovariance}}\xspace}

\def\MagUp {\mbox{\em Mag\kern -0.05em Up}\xspace}

\ifthenelse{\boolean{uprightparticles}}%
{

 \def\Pnu         {\ensuremath{\upnu}\xspace}

 \def\Ptau        {\ensuremath{\uptau}\xspace}

 \def\Ppsi        {\ensuremath{\uppsi}\xspace}

 \def\PDelta      {\ensuremath{\Delta}\xspace}                 
 \def\PXi         {\ensuremath{\Xi}\xspace}                 
 \def\PLambda     {\ensuremath{\Lambda}\xspace}                 
 \def\PSigma      {\ensuremath{\Sigma}\xspace}                 
 \def\POmega      {\ensuremath{\Omega}\xspace}                 
 \def\PUpsilon    {\ensuremath{\Upsilon}\xspace}

 \def\PB      {\ensuremath{\mathrm{B}}\xspace}                 
                  
 \def\PD      {\ensuremath{\mathrm{D}}\xspace}

 \def\PJ      {\ensuremath{\mathrm{J}}\xspace}                 
 \def\PK      {\ensuremath{\mathrm{K}}\xspace}

 \def\PW      {\ensuremath{\mathrm{W}}\xspace}

 \def\PZ      {\ensuremath{\mathrm{Z}}\xspace}                 
                  
 \def\Pb      {\ensuremath{\mathrm{b}}\xspace}                 
 \def\Pc      {\ensuremath{\mathrm{c}}\xspace}

 \def\Pi      {\ensuremath{\mathrm{i}}\xspace}

 \def\Ps      {\ensuremath{\mathrm{s}}\xspace}

 \def\thebaroffset{0.0em}
}
{

 \def\Pnu         {\ensuremath{\nu}\xspace}

 \def\Ptau        {\ensuremath{\tau}\xspace}

 \def\Ppsi        {\ensuremath{\psi}\xspace}                 
                  
 \mathchardef\PDelta="7101
 \mathchardef\PXi="7104
 \mathchardef\PLambda="7103
 \mathchardef\PSigma="7106
 \mathchardef\POmega="710A
 \mathchardef\PUpsilon="7107
                  
 \def\PB      {\ensuremath{B}\xspace}                 
                  
 \def\PD      {\ensuremath{D}\xspace}

 \def\PJ      {\ensuremath{J}\xspace}                 
 \def\PK      {\ensuremath{K}\xspace}

 \def\PW      {\ensuremath{W}\xspace}

 \def\PZ      {\ensuremath{Z}\xspace}                 
                  
 \def\Pb      {\ensuremath{b}\xspace}                 
 \def\Pc      {\ensuremath{c}\xspace}

 \def\Pi      {\ensuremath{i}\xspace}

 \def\Ps      {\ensuremath{s}\xspace}

 \def\thebaroffset{0.18em}
}
\newcommand{\offsetoverline}[2][\thebaroffset]{\kern #1\overline{\kern -#1 #2}}%

\makeatletter
\ifcase \@ptsize \relax
  \newcommand{\miniscule}{\@setfontsize\miniscule{4}{5}}
\or
  \newcommand{\miniscule}{\@setfontsize\miniscule{5}{6}}
\or
  \newcommand{\miniscule}{\@setfontsize\miniscule{5}{6}}
\fi
\makeatother

\DeclareRobustCommand{\optbar}[1]{\shortstack{{\miniscule (\rule[.5ex]{1.25em}{.18mm})}
  \\ [-.7ex] $#1$}}

\def\taup       {{\ensuremath{\Ptau^+}}\xspace}

\def\neu        {{\ensuremath{\Pnu}}\xspace}

\def\neut       {{\ensuremath{\neu_\tau}}\xspace}

\def\W      {{\ensuremath{\PW}}\xspace}

\def\Z      {{\ensuremath{\PZ}}\xspace}

\def\squark    {{\ensuremath{\Ps}}\xspace}

\def\cquark    {{\ensuremath{\Pc}}\xspace}

\def\bquark    {{\ensuremath{\Pb}}\xspace}

\def\KorKbar {\kern \thebaroffset\optbar{\kern -\thebaroffset \PK}{}\xspace}

\def\Dbar    {{\ensuremath{\offsetoverline{\PD}}}\xspace}

\def\DorDbar {\kern \thebaroffset\optbar{\kern -\thebaroffset \PD}\xspace}

\def\Dzb     {{\ensuremath{\Dbar{}^0}}\xspace}

\def\B       {{\ensuremath{\PB}}\xspace}

\def\BorBbar {\kern \thebaroffset\optbar{\kern -\thebaroffset \PB}\xspace}

\def\Bd      {{\ensuremath{\B^0}}\xspace}

\def\BdorBdbar {\kern \thebaroffset\optbar{\kern -\thebaroffset \Bd}\xspace}

\def\Bs      {{\ensuremath{\B^0_\squark}}\xspace}

\def\BsorBsbar {\kern \thebaroffset\optbar{\kern -\thebaroffset \Bs}\xspace}
\def\Bc      {{\ensuremath{\B_\cquark^+}}\xspace}

\def\jpsi     {{\ensuremath{{\PJ\mskip -3mu/\mskip -2mu\Ppsi\mskip 2mu}}}\xspace}

\def\Y#1S{\ensuremath{\PUpsilon{(#1S)}}\xspace}

\def\LorLbar     {\kern \thebaroffset\optbar{\kern -\thebaroffset \PLambda}\xspace}

\def\ra                 {\ensuremath{\rightarrow}\xspace}
\def\to                 {\ensuremath{\rightarrow}\xspace}

\newcommand{\mBc}{{\ensuremath{m_{\Bc}}}\xspace}

\def\Vcb  {{\ensuremath{V_{\cquark\bquark}}}\xspace}

\def\AT#1     {\ensuremath{A_{\mathrm{T}}^{#1}}\xspace}           

\def\C#1      {\ensuremath{\mathcal{C}_{#1}}\xspace}                       
\def\Cp#1     {\ensuremath{\mathcal{C}_{#1}^{'}}\xspace}                    
\def\Ceff#1   {\ensuremath{\mathcal{C}_{#1}^{\mathrm{(eff)}}}\xspace}        
\def\Cpeff#1  {\ensuremath{\mathcal{C}_{#1}^{'\mathrm{(eff)}}}\xspace}       
\def\Ope#1    {\ensuremath{\mathcal{O}_{#1}}\xspace}                       
\def\Opep#1   {\ensuremath{\mathcal{O}_{#1}^{'}}\xspace}                    

\newcommand{\aunit}[1]{\ensuremath{\text{\,#1}}}       

\newcommand{\tev}{\aunit{Te\kern -0.1em V}\xspace}
\newcommand{\gev}{\aunit{Ge\kern -0.1em V}\xspace}
\newcommand{\mev}{\aunit{Me\kern -0.1em V}\xspace}
\newcommand{\kev}{\aunit{ke\kern -0.1em V}\xspace}
\newcommand{\ev}{\aunit{e\kern -0.1em V}\xspace}
\newcommand{\mevc}{\ensuremath{\aunit{Me\kern -0.1em V\!/}c}\xspace}
\newcommand{\gevc}{\ensuremath{\aunit{Ge\kern -0.1em V\!/}c}\xspace}
\newcommand{\mevcc}{\ensuremath{\aunit{Me\kern -0.1em V\!/}c^2}\xspace}
\newcommand{\gevcc}{\ensuremath{\aunit{Ge\kern -0.1em V\!/}c^2}\xspace}

\def\gsim{{~\raise.15em\hbox{$>$}\kern-.85em
          \lower.35em\hbox{$\sim$}~}\xspace}
\def\lsim{{~\raise.15em\hbox{$<$}\kern-.85em
          \lower.35em\hbox{$\sim$}~}\xspace}

\def\evtgen     {\mbox{\textsc{EvtGen}}\xspace}

\def\photos     {\mbox{\textsc{Photos}}\xspace}

\def\pythia     {\mbox{\textsc{Pythia}}\xspace}

\def\tell1  {TELL1\xspace}
\def\ukl1   {UKL1\xspace}

\def\BcToTauNu{\Bc \ra \taup \neut }

\def \Vcb{\ensuremath{V_{cb}}\xspace}

\def \fBc {\ensuremath f_{B_{c}}\xspace}
\def \mBc {\ensuremath m_{B_{c}}\xspace}

\begin{document}

\maketitle

\begin{center}{\footnotesize \it
\noindent
$^{1}$ Universit\'e Paris-Saclay, CNRS/IN2P3, IJCLab, 91405 Orsay, France \\
$^{2}$ Universit\'e Paris-Saclay, 91400, Orsay, France \\
$^{3}$ European Organization for Nuclear Research (CERN), Geneva, Switzerland \\
$^{4}$ Institute of Physics, École Polytechnique Fédérale de Lausanne (EPFL), Lausanne, Switzerland}

\vspace{0.5cm}

{\footnotesize
{{Email:~}}{\bf\color{blue} yasmine.amhis@ijclab.in2p3.fr,
marie.hartmann@universite-paris-saclay.fr,
clement.helsens@cern.ch, 
donal.hill@cern.ch, 
olcyr.sumensari@ijclab.in2p3.fr}
}
\end{center}
\bigskip

\hrule
\begin{abstract}\noindent
This paper presents the prospects for a precise measurement of the branching fraction of the leptonic  $B_{c}^+\to \tau^+ \nu_\tau$ decay at the Future Circular Collider (FCC-ee) running at the $Z$-pole.  A detailed description of the simulation and analysis framework is provided. 
To select signal candidates, two Boosted Decision Tree algorithms are employed and optimised. The first stage suppresses inclusive $b\bar{b}$, $c\bar{c}$, and $q\bar{q}$ backgrounds using event-based topological information. A second stage utilises the properties of the hadronic $\tau^{+} \to \pi^+ \pi^+ \pi^- \bar{\nu}_\tau$ decay to further suppress these backgrounds, and is also found to achieve high rejection for the $\BTauNu$ background. The number of $B_{c}^+\to \tau^+ \nu_\tau$ candidates is estimated for various Tera-$Z$ scenarios, and the potential precision of signal yield and branching fraction measurements evaluated. The phenomenological impact of such measurements on various New Physics scenarios is also explored. 
\end{abstract}
\hrule




\section{Introduction }
\label{sec:introduction}
Leptonic pseudoscalar meson decays such as \BcToTauNu are theoretically clean probes to test for the presence of physics beyond the Standard Model (SM). The only hadronic inputs required to compute their decay branching fractions in the SM are the decay constants, which have been precisely determined for several transitions by means of numerical simulations of QCD on the lattice (LQCD)~\cite{Aoki:2019cca}.
In the past several years, numerous discrepancies from SM predictions have been observed in tree-level~\cite{Lees:2013uzd,Aaij:2015yra,Aaij:2017deq,Aaij:2017tyk,Huschle:2015rga,Hirose:2016wfn,Hirose:2017dxl,Belle:2019rba} and loop-induced~\cite{Aaij:2014ora,Aaij:2017vbb,Aaij:2019wad,Aaij:2019bzx} semileptonic $b$-hadron decays, often referred to as the $B$-physics anomalies. The \BcToTauNu decay\footnote{ Charge conjugation is implied throughout this work, unless stated otherwise.} can be directly related to the anomalies in tree-level decays as they occur through the same quark-level transition, $\bquark\to \cquark\tau \nu_\tau$, thus offering a clean and independent check of these experimental results~\cite{Alonso:2016oyd,Li:2016vvp}. Furthermore, \BcToTauNu decays are highly sensitive probes of pseudoscalar contributions from New Physics (NP), as predicted for instance in extensions of the SM Higgs sector, such as Two-Higgs-Doublet Models (2HDM)~\cite{Branco:2011iw}, as well as in specific leptoquark models~\cite{Buchmuller:1986zs,Dorsner:2016wpm}.

Despite the fairly large branching ratio of the $B_c^+ \to \tau^+ \nu_\tau$ decay within the SM ($\approx 2\%$), an observation and precise measurement of its properties are very challenging at a hadron collider. This is due to the large missing energy in the final state, no knowledge of the centre-of-mass energy of the $b\bar{b}$ production process, high backgrounds due to the hadronic environment with multiple primary vertices (PVs), and the lack of any reconstructible $B_c^+$ decay vertex. Moreover, these decays cannot be studied at $B$-factories since $B_c^+$ mesons are too heavy to be produced from $\Upsilon(5S) \to b\bar{b}$ decays. For these reasons, future $Z$-factories provide a unique environment to study these processes in the future. Indeed, the large number $Z$ bosons produced, up to $N_Z \approx 5\times 10^{12}$ in the case of FCC-ee (so-called Tera-$Z$), together with the possibility to constrain the missing energy from the neutrinos, would make this measurement possible.  In this paper, the feasibility of performing such a measurement at FCC-ee in $Z$-pole operation~\cite{Gomez-Ceballos:2013zzn, Mangano:2651294,Benedikt:2651299} is demonstrated, employing the $\TauThreePi$ mode to provide the $\tau^+$ decay vertex and thus a measure of the combined $B_c^+$ and $\tau^+$ flight distance. This reconstruction method offers additional means of background rejection, due to the lower lifetime of the $B_c^+$ meson relative to lighter $b$-hadrons, and the different resonant properties of hadronic $\tau^+$ decays compared to backgrounds from $b$- and $c$-hadrons. In the context of the CEPC project, a similar study has been presented in Ref.~\cite{Zheng:2021xuq}, where leptonic decays of the $\tau^+$ were considered.


The branching fraction of the \BcToTauNu decay in the SM can be written as
\begin{equation}
    \mathcal{B} (\BcToTauNu )^{\textrm {SM}} =\tau_{B_c} \frac{G_{F}^2 |\Vcb|^{2}\fBc^{2} \mBc m_\tau^2}{8\pi} \left( 1- \frac{m_\tau^{2}}{\mBc^{2}} \right )^{2}, 
\end{equation}
\sloppypar
\noindent where $G_F$ is the Fermi constant, $\mBc$ and $m_\tau$ are the the $B_c^+$ meson and $\tau^+$ lepton masses, respectively,  $\tau_{B_c}$ denotes the  $B_c^+$ meson lifetime \cite{Zheng:2021xuq},  $\Vcb$ is the CKM matrix element for $b \to c$ transitions~\cite{Amhis:2019ckw}, and $f_{B_c}$ represents the $B_c^+$ meson decay constant,
$\fBc=427(6)$~MeV, which has been computed via LQCD simulations in \cite{McNeile:2012qf} (see also Ref.~\cite{Colquhoun:2015oha}). By combining these inputs with the latest value of $|V_{cb}|^\mathrm{excl.}=39.09(68)\times 10^{-3}$~\cite{Aoki:2019cca}, determined from exclusive $B\to D^{(\ast)}\ell \nu_\ell$ decays using the BGL parameterization for the $B\to D^{(\ast)}$ form factors~\cite{Boyd:1994tt,Boyd:1997kz},
\begin{equation}
\mathcal{B}(B_c^+\to \tau^+ \nu_\tau)^\mathrm{SM}=1.95(9)\times 10^{-2}
\end{equation}
is obtained. One of the challenges of a precise measurement of the \BcToTauNu decay branching fraction is to properly normalise the measurement, in order to avoid relying on the unknown $B_c^+$ meson hadronisation fraction, $f(B_c^\pm) \equiv f(b\to B_c^\pm)$~\cite{Mangano:1997md}. This was the main caveat of previous attempts to extract limits on $\mathcal{B}(B_c^+ \to \tau^+ \nu_\tau)$ from LEP data~\cite{Akeroyd:2008ac,Akeroyd:2017mhr}, as discussed for example in Refs.~\cite{Blanke:2018yud,Blanke:2019qrx}. In this work, the possibility of normalising the measurement to the $B_c^+\to J/\psi \mu^+ \nu_\mu$ mode is proposed, using the $B_c^+ \to J/\psi$ form factors recently computed via LQCD in Ref.~\cite{Harrison:2020gvo,Harrison:2020nrv} to predict $\mathcal{B}(B_c^+ \to J/\psi \mu^+ \nu_\mu)$. The $B_c^+ \to \tau^+ \nu_\tau$ branching fraction is then determined as follows,
\begin{align}
\begin{split}
    \mathcal{B}(B_c^+ \to \tau^+ \nu_\tau) = \frac{N(B_c^+ \to \tau^+ \nu_\tau)}{N(B_c^+ \to J/\psi \mu^+\nu_\mu)} &\times \dfrac{\epsilon(B_c^+ \to J/\psi \mu^+\nu_\mu)}{\epsilon(B_c^+ \to \tau^+ \nu_\tau)}\\[0.35em]
    &\times \dfrac{\mathcal{B}(J/\psi\to\mu^{+}\mu^{-})}{\mathcal{B}(\tau^+\to 3\pi \bar{\nu}_\tau)}\times \mathcal{B}(B_c^+\to J/\psi \mu^+ \nu_\mu)\,,
\end{split}
\end{align}
where the number of signal and normalisation candidates, \mbox{$N(B_c^+ \to \tau^+ \nu_\tau)$} and \mbox{$N(B_c^+ \to J/\psi \mu^+\nu_\mu)$}, are both measured in data. Their corresponding total efficiencies \mbox{$\epsilon(B_c^+ \to \tau^+ \nu_\tau)$} and \mbox{$\epsilon(B_c^+ \to J/\psi \mu^+\nu_\mu)$} are estimated using numerical simulations. The branching fractions of the $\jpsi$ to a pair of muons and the $\tau^{+}$ hadronic decay to three pions are taken from Ref.~\cite{PDG}, and the $B_c^+ \to J/\psi \mu^+ \nu_\mu$ branching fraction can be accurately predicted by using the LQCD results for the form-factors to be $\mathcal{B}(B_c^+\to J/\psi \mu^+\nu_\mu)^\mathrm{SM} = 0.0135(11)$~\cite{Harrison:2020gvo,Harrison:2020nrv}, where the $|V_{cb}|^\mathrm{excl.}$ value quoted above is used.
Alternatively one could use the fully reconstructed $B_c^+ \to J/\psi \pi^+$ decay mode as a normalisation instead of the partially reconstructed $B_c^+ \to J/\psi \mu^+ \nu_\mu$ decay. However, the branching fraction of the $B_c^+ \to J/\psi \pi^+$ decay is measured to be around 5\% of the $B_c^+ \to J/\psi \mu^+ \nu_\mu$ branching fraction~\cite{Bc2JpsiMuNu_Bc2JpsiPi_LHCb}, and as such the uncertainty due to finite normalisation statistics would be considerably larger. In addition, no reliable SM prediction for $\mathcal{B}(B_c^+ \to J/\psi \pi^+)$ is yet available.

Taking $B_c^+ \to J/\psi \mu^+ \nu_\mu$ as the normalisation mode, the ratio
\begin{align}
R_c\equiv  \dfrac{\mathcal{B}(B_c^+\to\tau^+\nu_\tau)}{\mathcal{B}(B_c^+\to J/\psi \mu^+\nu_\mu)} \stackrel{\mathrm{SM}}{=} 1.45(11)
\end{align}
can be measured. This ratio also offers a sensitive probe of NP contributions, with the advantage of being independent of both the $B_c^+$ production rate and $|V_{cb}|$, as discussed in Refs.~\cite{Becirevic:2020rzi,Banelli:2018fnx,Fleischer:2021yjo} for example. The limiting factor on the precision of $R_c$ is the uncertainty on the $B_c^{+}\to J/\psi$ form-factors, which amount to a theory uncertainty of $\approx 7 \%$ on $R_c$. This source of uncertainty can be improved in the future with updated LQCD computations, the precision of which can be considerably improved by performing a dedicated angular analysis of $B_c^+\to (J/\psi \to \mu^+\mu^-)\mu^+ \nu_\mu$ decays with the upgraded LHCb detector or at FCC-ee. Such measurements would be particularly useful to constrain the form factors in the large-recoil region.

The remainder of this paper is organised as follows: a description of the experimental setup, software, and simulated samples is provided in Section~\ref{sec:experimental-setup}; a demonstration of the  multivariate selections used to select signal decays with high purity is provided in Section~\ref{sec:analysis}, along with estimates of the achievable signal yield and branching fraction precision as a function of $N_Z$; the phenomenological impact of this measurement on well-motivated NP scenarios is discussed in Section~\ref{sec:interpretation}. 

\ 

The work in this paper has been conducted after the publication of the FCC conceptual design reports~\cite{Mangano:2651294,Benedikt:2651299,Benedikt:2018csr} and after the European Strategy Update for Particle Physics released its recommendation to investigate the technical and financial feasibility of a future hadron collider at CERN with a centre-of-mass energy of at least 100 TeV and with an electron--positron Higgs and electroweak factory as a possible first stage. This article provides a detailed description of the key ingredients for physics analyses at FCC-ee.

\section{Experimental environment}
\label{sec:experimental-setup}

\subsection{FCC-ee}
The international Future Circular Collider (FCC) study aims at a design of $p$-$p$, $\rm e^{+}e^{-}$, and $e$-$p$ colliders to be built in a new 100~km tunnel in the Geneva region. 
The $\rm e^{+}e^{-}$ collider (FCC-ee) has a centre of mass energy range between 91 (\Z-pole) and 375~GeV ($t\bar{t}$). The FCC-ee offers unprecedented possibilities for measuring the properties of the four heaviest particles of the SM (the Higgs, \Z, and \W bosons, and the top quark), but also those of the \bquark and \cquark quarks and of the $\tau$ lepton. In addition, circular colliders have the advantage of delivering collisions to multiple interaction regions, which allow different detector designs to be studied and optimised -- up to four are under consideration for FCC-ee. Moreover, the huge statistics anticipated at the \Z{} peak (the so-called ``Tera-\Z{}" run) brings specific challenges, as the systematic uncertainties of the measurements should be commensurate with their small statistical uncertainties.

\subsection{Simulation of the detector response}
The detector response has been simulated via the \delphes{}~software package~\cite{deFavereau:2013fsa}. It is a C++ framework, performing a fast multipurpose detector response simulation. The simulation includes a tracking system embedded in a magnetic field, calorimeters, and a muon system. The framework is interfaced to standard file formats (e.g.~Les Houches Event File or HepMC) and outputs observables such reconstructed charged tracks which can be used for dedicated analyses. The simulation of the detector response takes into account the effect of the magnetic field, the granularity of the calorimeters, and sub-detector resolutions. In the pre-release 3.4.3pre10 used for this analysis, \delphes{} provides parameterised track information with the full covariance matrix using the \fasttrack software. 

The detector configuration considered is the Innovative Detector for Electron–positron Accelerators (IDEA) concept. It comprises a silicon pixel vertex detector, a large-volume extremely-light short-drift wire chamber surrounded by a layer of silicon micro-strip detectors, a thin, low-mass superconducting solenoid coil, a pre-shower detector, a dual-readout calorimeter, and muon chambers within the magnet return yoke~\cite{Benedikt:2651299}. The \delphes{}~configuration card used for this analysis is accessible in the repository given in Ref.~\cite{helsens_clement_2021_4817845}. Finally, the \kSimDelphes~\cite{valentin_volkl_2021_4748578} project converts \delphes{} objects to \edmhep{}~\cite{valentin_volkl_2021_4785063}, and the subsequent Monte Carlo (MC) production is performed in the common \edmhep{} data format.

\subsection{Monte-Carlo Production}
MC event samples are used to simulate the response of the detector to signal and background processes. Signal and background events are generated with \pythia~\cite{Sjostrand:2014zea} version 8.303 using the leading order cross-section from the generator with no K-factor. Decays of unstable particles are described using \evtgen~\cite{Lange:2001uf} version 02.00.00, in which final-state radiation is generated using \photos~\cite{Davidson:2010ew}. Several important parameters are configured to be the same for all samples in the \pythia steering cards. At FCC-ee~\cite{Benedikt:2651299}, the energy of the beams is distributed according to a Gaussian function. At the \Z{} peak, the beam energy spread amounts to 0.132\% of the incoming beam energy, half the \Z{} boson mass, which equates to 0.0602\,GeV. 
The position of the interaction region depends on the running conditions of the machine. At the \Z{}-pole the bunch length is $\sigma_z = 12.1$\,mm. The bunch dimensions in the transverse plane, at the interaction point (IP), are given by $\sigma_{x,y} = \sqrt{\beta^*_{x,y}\times \epsilon_{x,y}}$, where the values of the $\beta$ function at the IP, and the horizontal and vertical emittance $\epsilon_{x,y}$ are given in the FCC-ee Conceptual Design Report~\cite{Benedikt:2651299} and are approximately $\sigma_x=6.4$\,$\mu$m and $\sigma_y=28.3$\,nm. For Gaussian bunches, the PV distribution in $(x, y, z)$ is well approximated by a 3--dimensional Gaussian distribution, with
\begingroup
\allowdisplaybreaks
\begin{align*}
    &\sigma_x^{PV} = 1\bigg{/}\sqrt{2\left(\frac{\mathrm{cos}^2\alpha}{\sigma_x} + \frac{\mathrm{sin}^2\alpha}{\sigma_z}\right)} \simeq \frac{\sigma_x}{\sqrt{2}}\,,\\
    &\sigma_y^{PV} = \frac{\sigma_y}{\sqrt{2}}\,,\\
    &\sigma_z^{PV} =  1\bigg{/}\sqrt{2\left(\frac{\mathrm{cos}^2\alpha}{\sigma_z} + \frac{\mathrm{sin}^2\alpha}{\sigma_x}\right)}\,,
\end{align*}
\endgroup
where $\alpha$ denotes the half-crossing angle of 15\,mrad. This yields values for the primary vertex smearing, $\sigma_x^{PV} = 4.5$\,$\mu$m, $\sigma_y^{PV} = 20$\,nm, and $\sigma_z^{PV} = 0.3$\,mm. 
    
All of the generator configurations and steering cards are documented and preserved~\cite{helsens_clement_2021_4817845}. The production of Monte-Carlo events is achieved using the FCC common tools and CERN computing and storage resources~\cite{helsens_clement_2021_4817856}. Approximately $10^{10}$ events are produced, representing about 55\,TB of disk space. Dedicated productions with orthogonal seeds between the analysis and multivariate training samples have been considered in order to avoid over--training.

\subsection{Analysis framework}
A sophisticated analysis framework has been developed for all FCC analyses using the common \edmhep{} data format. It is based on RDataFrames~\cite{enrico_guiraud_2017_260230}, where C++ code is conveniently compiled in a \textsc{ROOT}~\cite{rene_brun_2019_3895860} dictionary as ``analysers" which are subsequently called in Python~\cite{helsens_clement_2021_4817870}. Several external packages such as \textsc{ACTS}~\cite{xiaocong_ai_2020_3937454}, \textsc{FastJet}~\cite{Cacciari:2011ma}, and \textsc{awkward}~\cite{jim_pivarski_2021_4746371} are included. The analysis code is distributed via the CERN virtual machine file system cvmfs, and can be run locally or on batch systems. The complete software stack used to produce the results in this paper can be accessed and the results reproduced~\cite{key4hep_collaboration_2021_4818082}.

\subsection{Specificity for this analysis}
For this analysis, dedicated new features of the analysis framework have been developed and are explained in this section. 

\begin{figure}[h!]
\centering
\includegraphics[width = 0.49\textwidth]{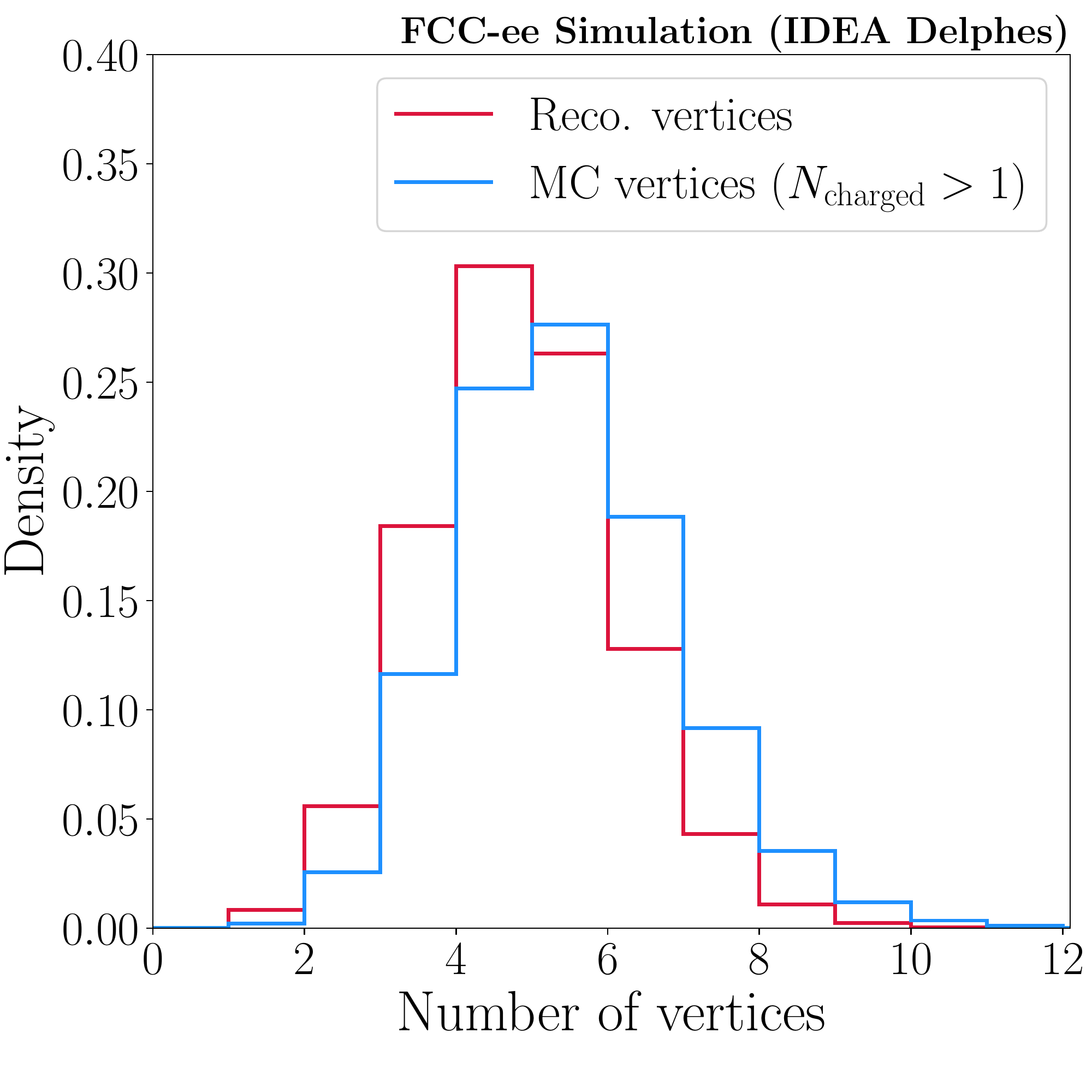} \put(-165,180){(a)}
\includegraphics[width = 0.49\textwidth]{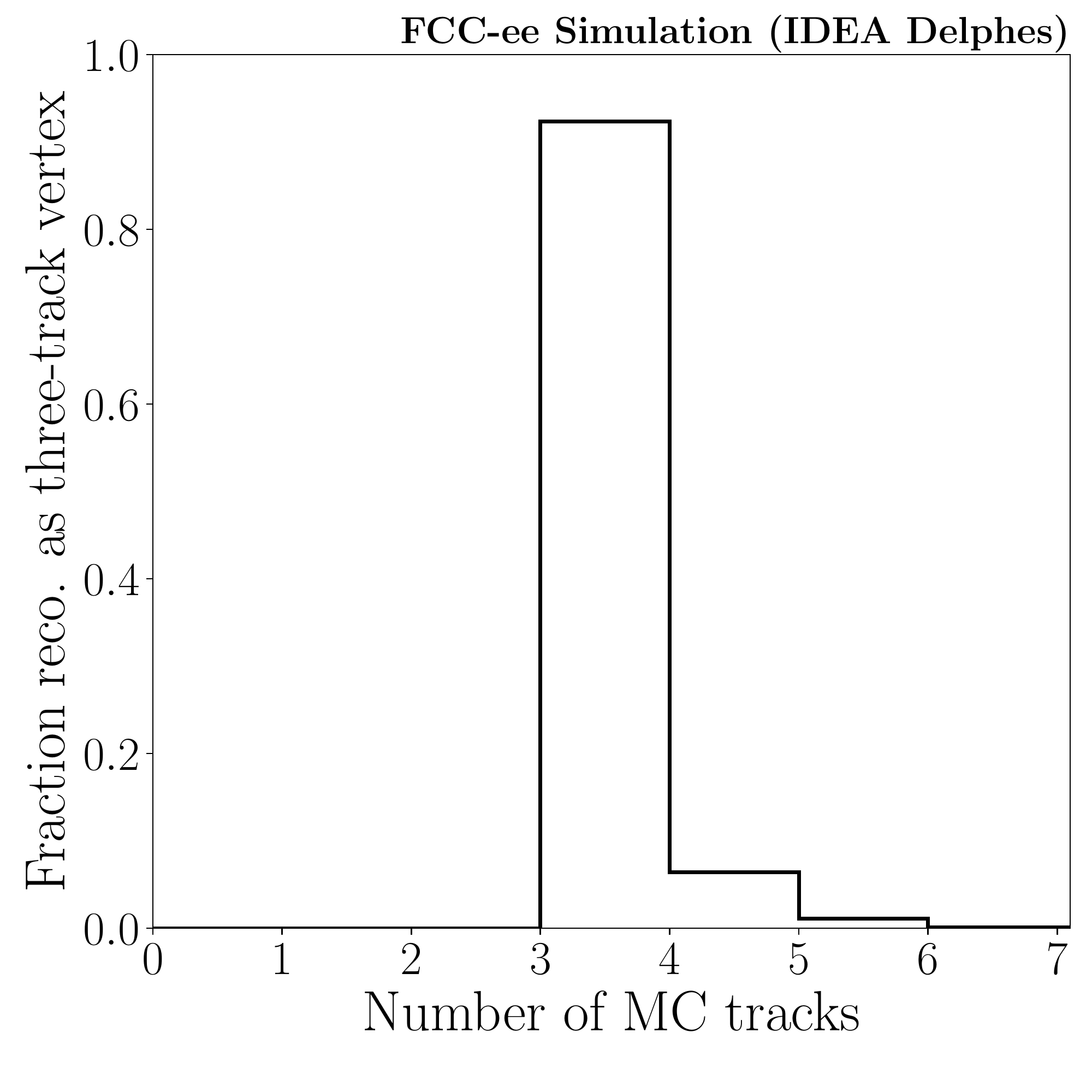}\put(-165,180){(b)}
\caption{(a) Comparison of normalised number truth-level MC vertices (red) and number of reconstructed vertices (blue), where MC vertices with one or more charged particle are shown. (b) Proportion of three-track reconstructed vertices that correspond to MC vertices with different numbers of charged particles. Over 90\% of reconstructed three-track vertices originate from a true three-track MC vertex.}
\label{fig:vertexing}
\end{figure}

\textbf{Perfect vertex seeding:}
Excellent vertex finding is crucial for this analysis. While detailed investigations are ongoing to estimate the impact of imperfect vertexing, for the following results it is assumed that vertices can be perfectly seeded. The procedure is to first find all of the MC vertices by selecting stable charged particles originating from the same point. From those MC vertices, reconstruction-level vertices are fitted using the reconstructed tracks associated to the MC particles attached to the MC vertex. A plot comparing the MC and reconstruction-level number of vertices is shown in Fig.~\ref{fig:vertexing} (a). This procedure properly takes into account the migration of higher number of MC tracks to a given reconstructed multiplicity, as illustrated in Fig.~\ref{fig:vertexing} (b), where about 7\% of the three-track reconstructed vertices originate from a four-track MC vertex.

\textbf{Perfect particle identification:}
In the energy range considered for the identification of pions and kaons in the analysis ($\sim10$\,GeV), extremely good discrimination is expected, thus for this measurement it is considered that the pions and kaons can be perfectly identified.

Processing the first stage of the analysis over the full sample statistics, and calculating  complex quantities (such as the thrust described in Section~\ref{subsection:thrust}, vertexing, building the candidates), takes approximately half a day on a batch system. The total sample size after first-stage processing is $\sim280$\,GB, representing a reduction factor of 200. The second stage of the analysis can then be run locally very quickly, as can all of the final analysis steps described in Section~\ref{sec:analysis}. 
\section{Analysis}
\label{sec:analysis}

To demonstrate the feasibility of a measurement of the $\BcTauNu$ mode at FCC-ee, a selection procedure based on the differences in reconstructed event and candidate properties in simulated $\BcTauNu$ and inclusive hadronic $Z$ decays is developed. The selection consists of a series of rectangular cuts in addition to two boosted decision tree (BDT) classifiers, which together achieve a high purity final dataset with a clearly identifiable signal component. The selection is now described, and estimates for the signal yield precision derived via template fits to combined samples of signal and background decays. A discussion of the potential branching ratio precision as a function of $N_Z$ is also given. All the tools used for the following analysis are available in Ref.~\cite{perez_emmanuel_2021_4818652}.

\subsection{Signal and background samples}
\label{sec:sig_bkg_samples}

All samples used in the selection studies are generated and processed using the framework detailed in Sec.~\ref{sec:experimental-setup}. For the BDT training, orthogonal samples of signal and background events are generated; these samples are not used in any other analysis steps to avoid biasing the BDT distributions.

Simulated samples of $\BcTauNu$ with $\TauThreePi$ decays are considered as signal in the selection studies. The samples are generated using \evtgen, where the $B_c^+$ is decayed using the \texttt{SLN} model and the $\TauThreePi$ decay is generated using the \texttt{TAUHADNU} model. The \texttt{TAUHADNU} model is used instead of alternative models such as \texttt{TAUOLA} to enable highly efficient truth-matching of the pions to a $\tau^+$ parent; the difference in $\tau^+$ decay product kinematics across models is not sufficient to alter the outcome of the selection studies.  

For both BDT training stages, inclusive samples of $Z \to b\bar{b}$, $c\bar{c}$, and $q\bar{q}$ decays are used as background, where $q \in \{u,d,s\}$. The samples are generated using Pythia, and are found to have consistent distributions when compared to inclusive samples generated using \textsc{EvtGen}. The background samples are combined according to known hadronic $Z$ branching fractions~\cite{PDG} and the total efficiencies of selection cuts applied prior to the training steps. 

Prior to the BDT cut optimisation, none of the $10^9$ inclusive $Z \to c\bar{c}$ and $Z \to q\bar{q}$ events are found to pass sufficiently tight cuts on both BDTs. As such, background from these sources is not considered in the optimisation or subsequent fit studies. After the same cuts, the remaining statistics in the inclusive $Z \to b\bar{b}$ sample are found to be insufficient for determining the background rejection accurately in the cut optimisation. To boost the background statistics for the optimisation, samples of exclusive $b$-hadron decays are generated, where the decay modes are chosen based on the composition of the remaining inclusive $Z \to b \bar{b}$ sample. The following decays are considered:
\begin{itemize}
    \item $B \to D \tau^+ \nu_\tau$
    \item $B \to D^* \tau^+ \nu_\tau$
    \item $B \to D \pi^+ \pi^+ \pi^-$
    \item $B \to D^* \pi^+ \pi^+ \pi^-$
    \item $B \to D D_s^+$
    \item $B \to D^* D_s^+$
    \item $B \to D^* D_s^{*+}$
\end{itemize}
where $B \in \{B^0, B^+, B_s^0, \Lambda_b^0\}$ and the corresponding $D \in \{D^-, \Dzb, D_s^-, \Lambda_c^-\}$. In each of the exclusive $b$-hadron samples, all of the $b$-hadron decay products are decayed inclusively. The list of exclusive decays considered is not exhaustive, and covers around 10\% of the decay width for each $B$ hadron. As a result, a factor 2.5 difference in rate relative to the inclusive $Z \to b\bar{b}$ sample is observed after tight BDT cuts. This factor is used to scale the exclusive sample yield estimates in the optimisation procedure, in order to avoid underestimating the expected background level.

\subsection{Thrust axis and event hemisphere definitions}
\label{subsection:thrust}
The signal selection relies on the large missing energy signature of $\BcTauNu$ decays, which arises due to the presence of two neutrinos in the final state. In a $Z \to b\bar{b}$ event involving a signal decay, the signal side of the event will on average contain considerably more missing energy than the non-signal side. This is in contrast to a general $Z \to b\bar{b}$ event, where both sides of the event will contain similar amounts of reconstructed energy on average. To determine the energy imbalance in an event, it is necessary to divide the event into two hemispheres, each corresponding to one of $b$-quarks produced in the $Z$ decay. 

In this analysis, hemispheres are defined event-by-event using the plane normal to the thrust axis. The thrust is the unit vector $\hat{\textbf{n}}$ which minimises
\begin{equation}
    T = \frac{\sum_{i}|\textbf{p}_i \cdot \hat{\textbf{n}}|}{\sum_i |\textbf{p}_i|},
\end{equation}
where $\textbf{p}_i$ is the momentum vector of the i$^\text{th}$ reconstructed particle. The axis along which $\hat{\textbf{n}}$ lies is referred to as the thrust axis, and provides a measure of the direction of the quark pair produced in the $Z$ decay. Reconstructed particles are assigned to either hemisphere based on the angle $\theta$ between their momentum vector and the thrust axis. In this analysis, the thrust is defined to point towards the hemisphere with less total energy; for a particle in the minimum energy hemisphere, $\cos(\theta) \geq 0$, while particles in the maximum energy hemisphere have $\cos(\theta) < 0$. 

\subsection{First-stage BDT}

The first step of the selection is designed to separate signal and background decays based on the energy signatures of both hemispheres and other general properties of the event. Prior to the first-stage BDT, events are required to contain a reconstructed PV and at least one reconstructed $3\pi$ candidate with an associated vertex. In addition, at least one of the $3\pi$ candidates is required to reside in the minimum energy hemisphere, since this hemisphere is more likely to contain the signal decay.

\begin{figure}[h!]
\centering
\includegraphics[width = 0.49\textwidth]{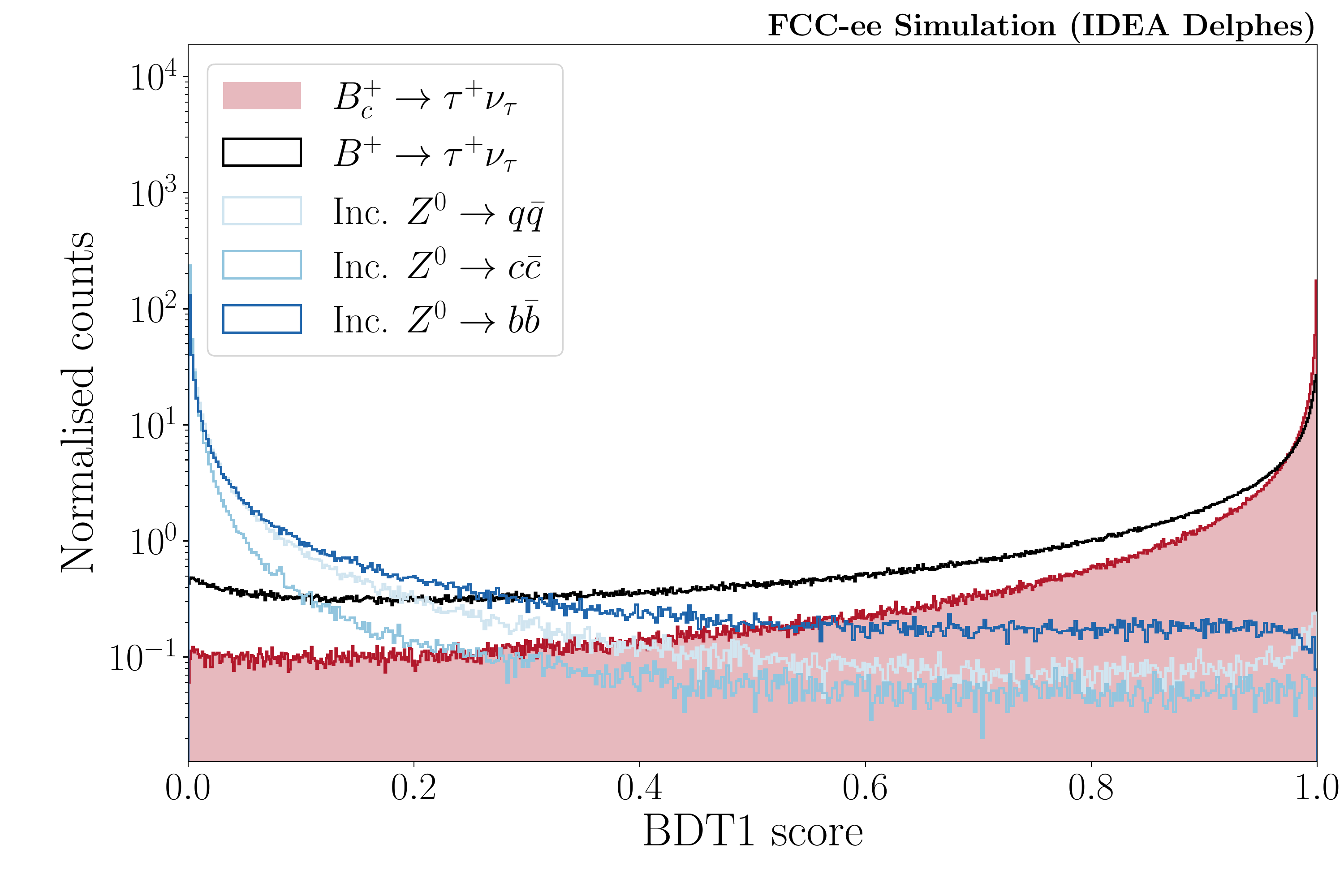} \put(-30,120){(a)}
\includegraphics[width = 0.49\textwidth]{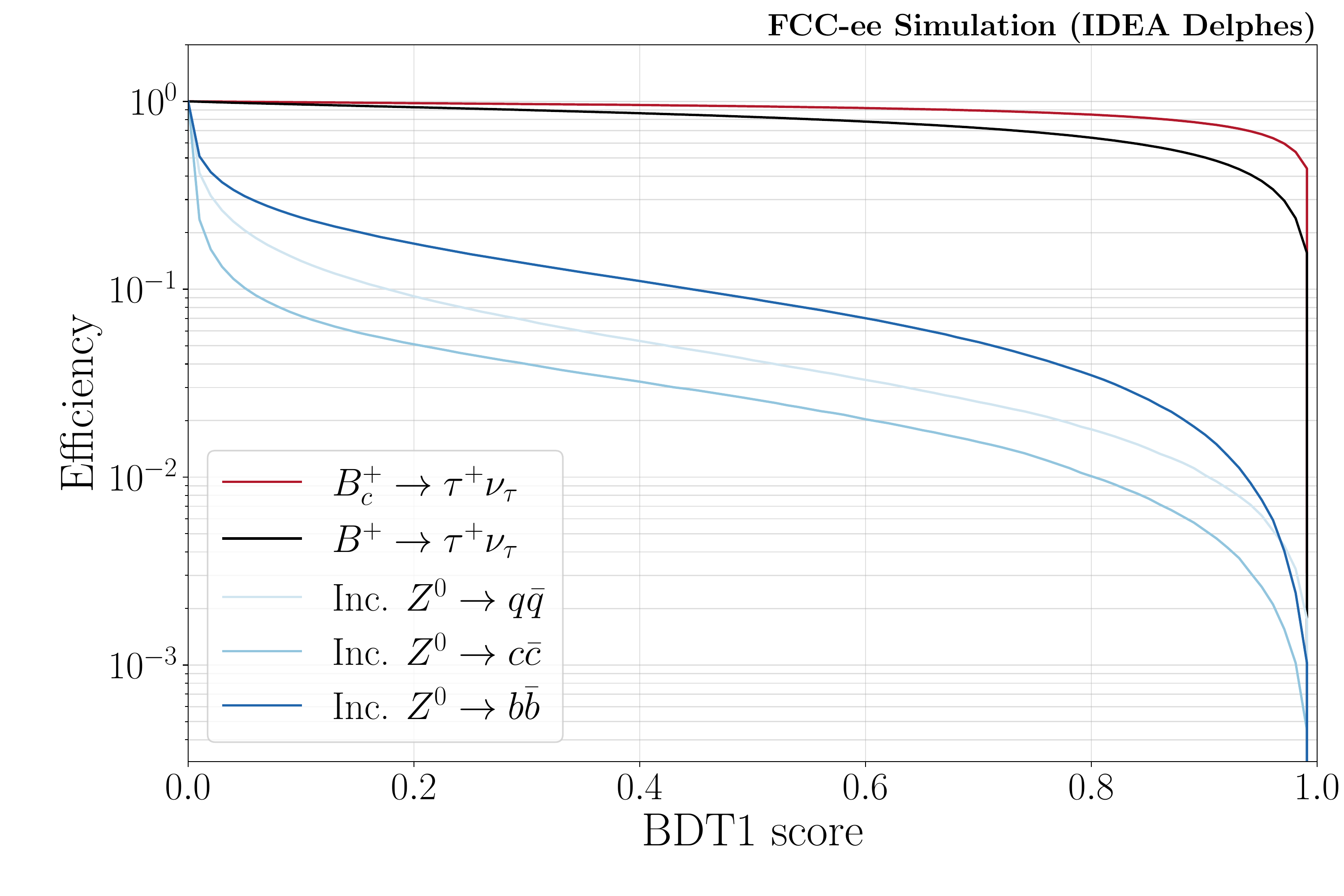}\put(-30,30){(b)}
\caption{(a) First-stage BDT distribution in signal, $\BTauNu$ background, and inclusive $Z$ background. (b) Efficiency of the first-stage BDT as a function of cut value.}
\label{fig:BDT1}
\end{figure}

The first-stage BDT is trained using \textsc{xgboost}~\cite{xgb} with a sample of $7\times 10^5$ signal events passing the above pre-selection and a combined sample of one million inclusive $Z \to b\bar{b}$, $c\bar{c}$, and $q\bar{q}$ events. The $Z \to b\bar{b}$ background sample is filtered to remove $\BcTauNu$ and $\BTauNu$ events. The relative proportion of each $Z$ decay type in the background sample is determined using their known hadronic $Z$ branching fractions~\cite{PDG}, multiplied by pre-selection efficiencies determined by the ratio of the number of events passing the pre-selection cuts relative to the number of generated events. The BDT is trained using the following features:
\begin{itemize}
    \item Total reconstructed energy in each hemisphere;
    \item Total charged and neutral reconstructed energies in each hemisphere;
    \item Charged and neutral particle multiplicities in each hemisphere;
    \item Number of tracks in the reconstructed PV;
    \item Number of reconstructed $3\pi$ candidates in the event;
    \item Number of reconstructed vertices in each hemisphere;
    \item Minimum, maximum, and average radial distance of all decay vertices from the PV.
\end{itemize}
The performance of the BDT is illustrated in Fig.~\ref{fig:BDT1}, where the BDT distributions in signal, $\BTauNu$, and each category of inclusive $Z$ background are shown alongside their corresponding efficiency profiles. The BDT is found to have a ROC curve area of 0.984, highlighting the excellent rejection of inclusive background achieved. Although $\BTauNu$ decays are not considered in the training background, the BDT achieves some rejection of this mode relative to signal. This is due to the different event-level properties of $\BTauNu$ and $\BcTauNu$ decays, which arise since the $B_c^+$ meson is produced with an associated charm quark that results in production of an associated charm hadron. Due to the finite charm hadron lifetime, this results in less energy and fewer tracks at the primary vertex on average compared to a $\BTauNu$ event, as well as more reconstructed displaced vertices. 

\FloatBarrier

\subsection{Second-stage BDT}

The second stage of the selection focuses on properties of the reconstructed $3\pi$ candidate and properties of other reconstructed decay vertices in the event. Prior to training the second-stage BDT, events are required to pass a cut of $> 0.6$ on the first stage BDT. This cut is over 90\% efficient on signal, and removes more than 90\% of all background types. As shown in Fig.~\ref{fig:BDT1}, the $Z \to b\bar{b}$ background is rejected least by the first-stage BDT, since it predominantly involves both $b \to c W$ and $c \to s W$ quark transitions, leading to more missing energy in the case of leptonic $W$ decays. In addition to the first-stage BDT cut, the difference in energy between the maximum and minimum energy hemispheres is required to exceed 10 GeV/$c^2$, in order to retain more signal-like events with a large energy imbalance.

In each event, a single $3\pi$ candidate is chosen as the signal candidate. The signal candidate must reside in the minimum energy hemisphere, and must have the smallest vertex fit $\chi^2$ of all $3\pi$ candidates in that hemisphere. Selected $3\pi$ candidates are required to have an invariant mass below that of the $\tau$ lepton, and must have at least one $m(\pi^+\pi^-)$ combination within the range $0.6-1.0$ GeV/$c^2$. These cuts retains candidates consistent with the $a_1(1260)^+ \to (\rho^0 \to \pi^+ \pi^-) \pi^+$ decay, via which all $\TauThreePi$ decays proceed.

The second-stage BDT is also trained using \textsc{xgboost}, with a sample of $5\times 10^5$ signal events and a combined sample of one million inclusive $Z$ decays. The proportions of $Z \to b\bar{b}$, $c\bar{c}$, and $q\bar{q}$ decays in the combined sample are determined using their known $Z$ branching fractions and the efficiencies of all pre-selection cuts.

\begin{figure}[h!]
\centering
\includegraphics[width = 0.49\textwidth]{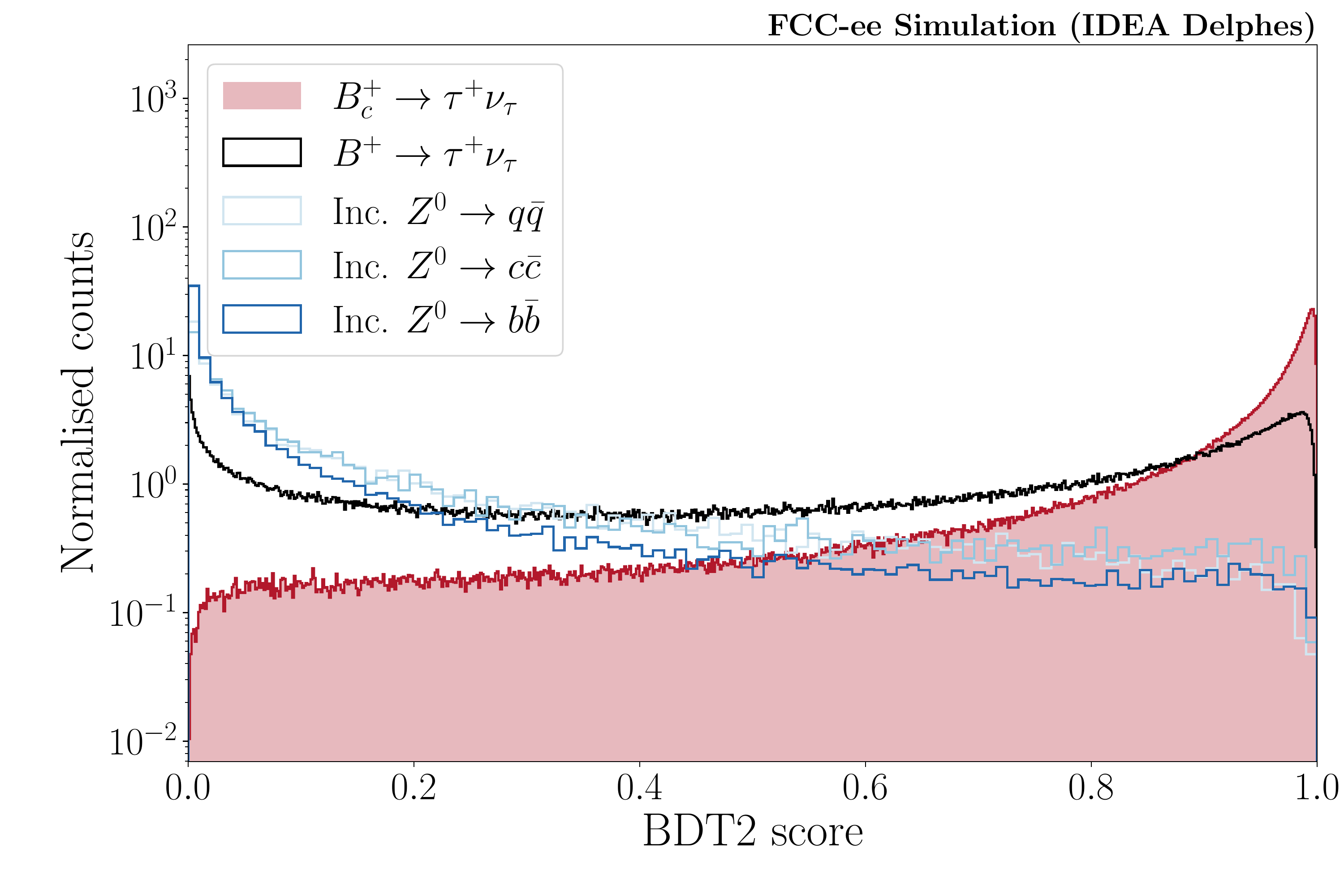}\put(-30,120){(a)}
\includegraphics[width = 0.49\textwidth]{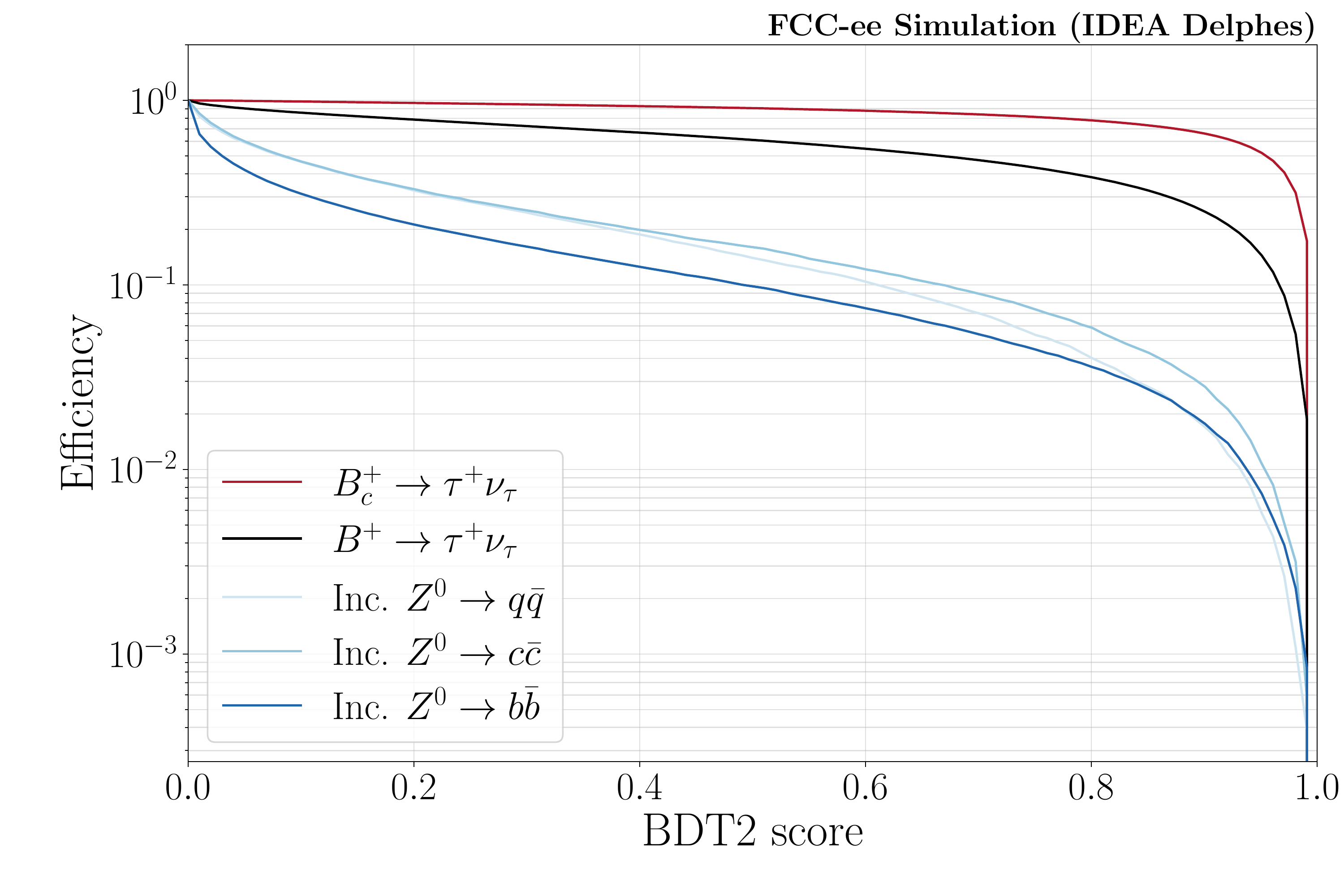}\put(-30,30){(b)}
\caption{(Left) Second-stage BDT distribution in signal, $\BTauNu$ background, and inclusive $Z$ background. (Right) Efficiency of the second-stage BDT as a function of cut value.}
\label{fig:BDT2}
\end{figure}

The BDT is trained on the following features:
\begin{itemize}
    \item $3\pi$ candidate mass, and masses of the two $\pi^+\pi^-$ combinations;
    \item Number of $3\pi$ candidates in the event;
    \item Radial distance of the $3\pi$ candidate from the PV;
    \item Vertex $\chi^2$ of the $3\pi$ candidate;
    \item Momentum magnitude, momentum components, and impact parameter (transverse and longitudinal) of the $3\pi$ candidate;
    \item Angle between the $3\pi$ candidate and the thrust axis;
    \item Minimum, maximum, and average impact parameter (longitudinal and transverse) of all other reconstructed decay vertices in the event;
    \item Mass of the PV;
    \item Nominal $B$ energy, defined as the $Z$ mass minus all reconstructed energy apart from the $3\pi$ candidate.
\end{itemize}
The performance of the BDT is illustrated in Fig.~\ref{fig:BDT2}, where the BDT distributions in signal, $\BTauNu$, and each category of inclusive $Z$ background are shown alongside their corresponding efficiency profiles. The BDT is found to have a ROC curve area of 0.966, indicating the high rejection of background achieved even after the BDT1 $> 0.6$ cut. The BDT is also found to reject $\BTauNu$ decays to a high level, owing to the larger lifetime of the $B^+$ meson compared to the $B_c^+$ meson which results in a greater $3\pi$ displacement from the PV on average. In addition, the lack of an associated charm hadron in $\BTauNu$ decays is also discriminated against by the second-stage BDT. To select a high-purity sample of signal decays, an optimisation procedure is employed to tune the two BDT cuts, using estimates for the signal and background yields expected at FCC-ee.

\subsection{Selection optimisation}

To determine the best BDT cut values, a two-dimensional optimisation procedure is performed. Prior to the optimisation, cuts of BDT1 $> 0.99$ and BDT2 $> 0.99$ are applied in order to focus on the signal region. A grid of 50 cuts for both BDT1 and BDT2 between 0.99 and 1.0 is scanned (2500 points in total), and at each point the expected signal ($S$) and background ($B$) yields are estimated. The point where the signal purity $P = S/(S+B)$ is maximised is taken to represent the best cut values for BDT1 and BDT2.

To estimate the signal yield at a given pair of BDT cuts, the following formula is used:
\begin{equation}
    N(B_c^+ \to \tau^+ \nu_\tau) = N_Z \times \mathcal{B}(Z \to b\bar{b}) \times 2 \times f(B_c^+) \times \mathcal{B}(B_c^+ \to \tau^+ \nu_\tau) \times \mathcal{B}(\tau^+ \to \pi^+ \pi^+ \pi^- \bar{\nu}_\tau) \times \epsilon,
\end{equation}
where $N_Z$ is the number of $Z$ bosons produced at FCC-ee, $\mathcal{B}(Z \to b\bar{b})$ is the known branching fraction of the $Z \to b\bar{b}$ decay~\cite{PDG}, the factor of two accounts for the fact that either $b$-quark from the $Z$ decay can produce a $B_c^+$ meson, $f(B_c^+) = 0.04\%$ is the $B_c^+$ meson production fraction taken from Pythia, $\mathcal{B}(B_c^+ \to \tau^+ \nu_\tau) = 1.94\%$ is the SM prediction for the signal decay branching fraction, $\mathcal{B}(\tau^+ \to \pi^+ \pi^+ \pi^- \bar{\nu}_\tau)$ is the known $\TauThreePi$ branching fraction~\cite{PDG}, and $\epsilon$ is the efficiency of the full selection at a given pair of BDT cuts determined from simulation.

With sufficiently tight cuts to BDT1 and BDT2, all $10^9$ generated inclusive $Z \to c\bar{c}$ and $Z \to q\bar{q}$ decays are rejected. As such, background from these sources is not considered in the optimisation routine. Insufficient statistics remain at tight BDT cuts in the inclusive $Z \to b\bar{b}$ sample to measure accurate background rejection figures. As such, the exclusive background samples described in Sec.~\ref{sec:sig_bkg_samples} are used to represent the remaining sources of background. The yield for a particular exclusive background decay $B \to D X$ at a particular pair of BDT cuts is given by
\begin{equation}
    N(B \to DX) = N_Z \times \mathcal{B}(Z \to b\bar{b}) \times 2 \times f(B) \times \mathcal{B}(B \to D X) \times \epsilon,
\end{equation}
where $f(B)$ is the hadron production fraction for a particular $b$-hadron taken from Pythia ($f(B^0) = 0.43$, $f(B^+) = 0.43$, $f(B_s^0) = 0.096$, $f(\Lambda_b^0) = 0.037$), $\mathcal{B}(B \to D X)$ is the decay mode branching fraction taken from Ref.~\cite{PDG} where measured, and $\epsilon$ is the efficiency determined using simulation. Where background mode branching fractions are not yet measured, assumptions are made based on the measured branching fractions of the most topologically similar decay modes. The total background level is given by a sum over all exclusive modes considered, with a multiplicative factor of 2.5 to account for the observed difference in rate between the inclusive $Z \to b\bar{b}$ and exclusive $b$-hadron samples.

\begin{figure}[h!]
\centering
\includegraphics[width = 0.4\textwidth]{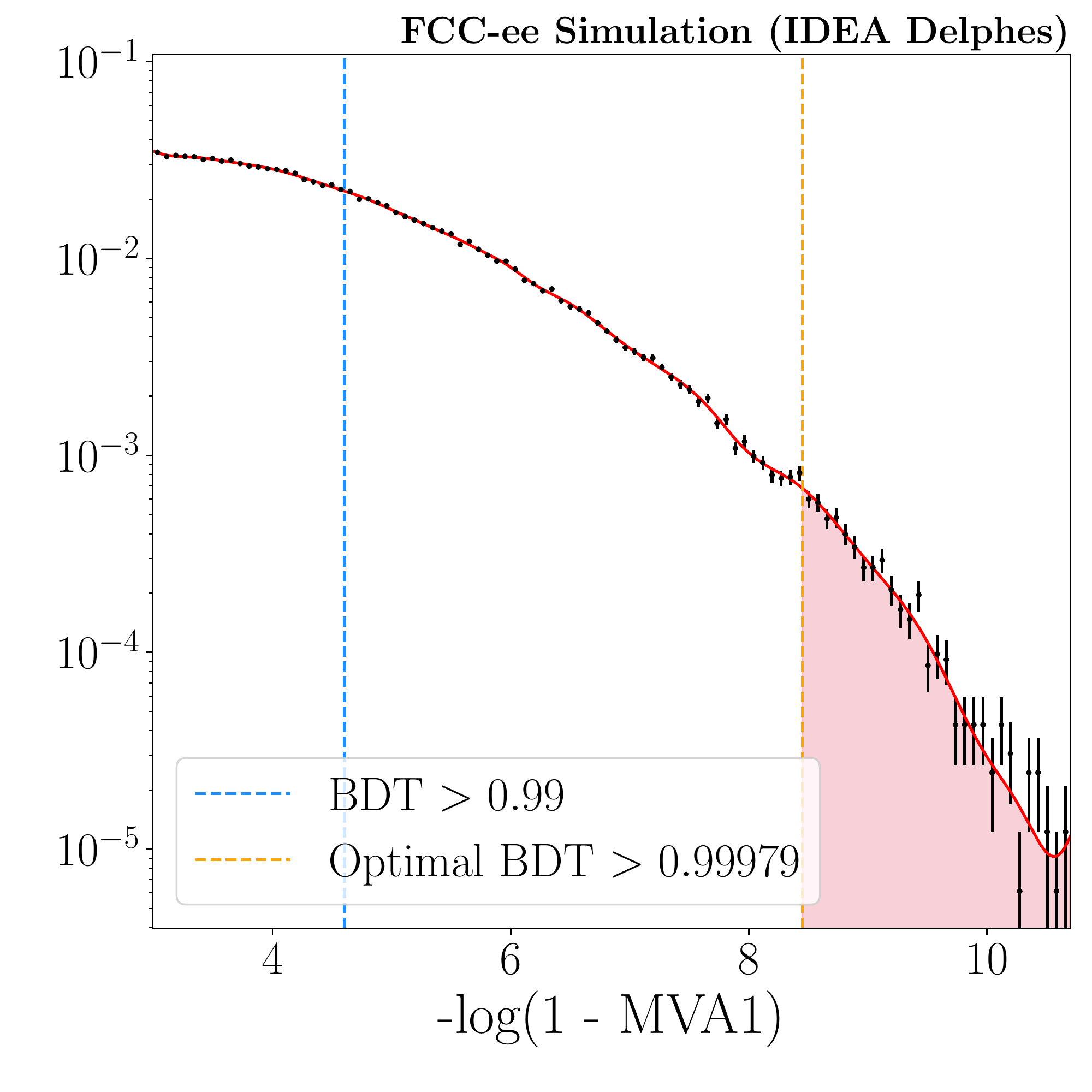}\put(-30,150){(a)}
\includegraphics[width = 0.4\textwidth]{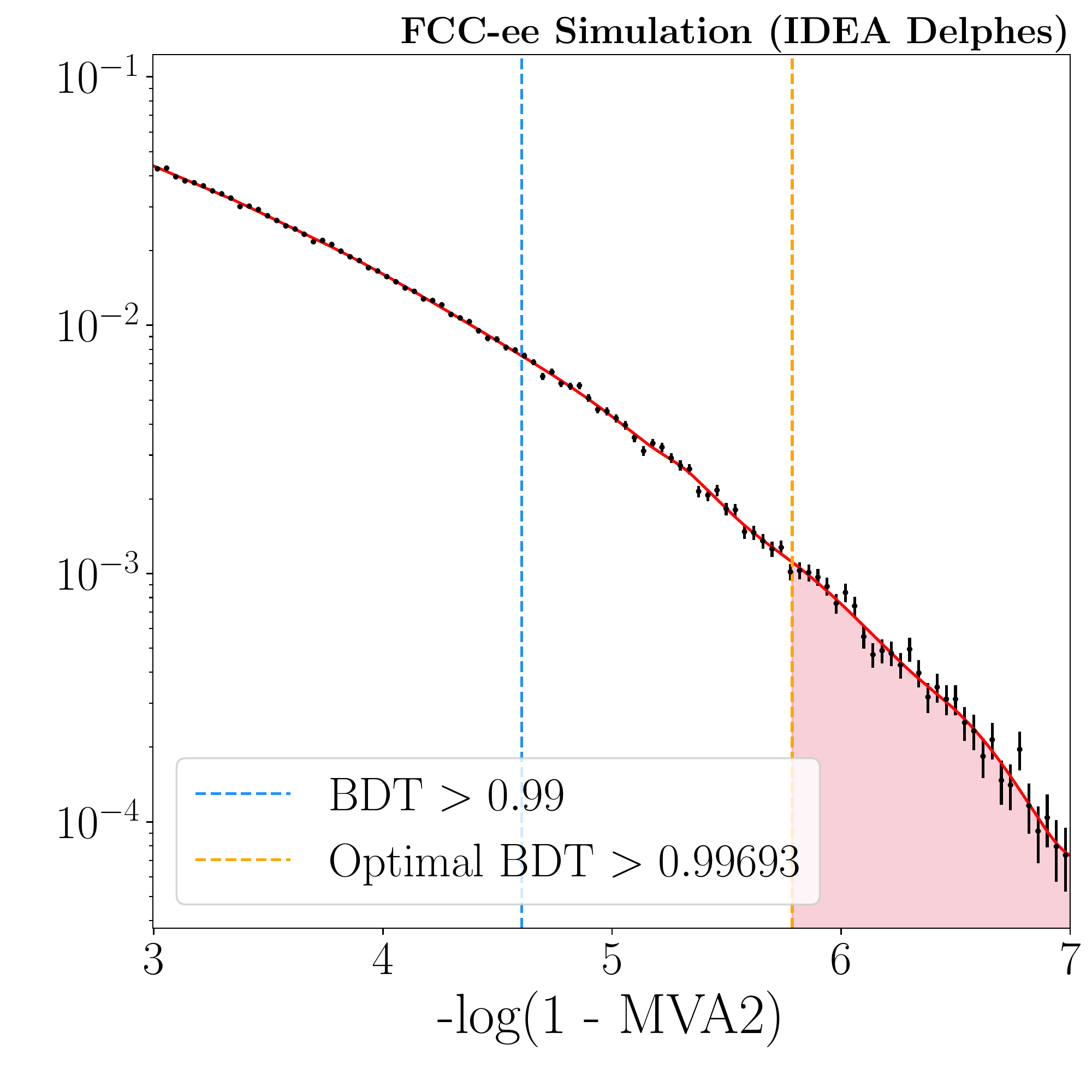}\put(-30,150){(b)}
\caption{(a) BDT1 distribution above 0.95 for a combined sample of exclusive $b$-hadron decays. (b) BDT1 distribution above 0.95 for a combined sample of exclusive $b$-hadron decays. The cubic spline parameterisations are shown by solid red lines, example cuts of BDT $> 0.99$ by the dashed blue lines, and the optimal BDT cuts by the dashed orange lines. The background efficiency given prior cuts of $> 0.95$ on both BDTs is given by the product of the spline integrals above the optimal cuts (red areas), where each integral is normalised to the respective spline integral across the full range.}
\label{fig:bkg_BDT_splines}
\end{figure}

To determine the background efficiencies, per-mode efficiencies for a combined BDT1 $> 0.95$ and \mbox{BDT2 $>0.95$} cut are first determined, as sufficient statistics remain in each exclusive sample to measure these efficiencies accurately. The efficiencies for subsequent BDT1 and BDT2 cuts relative to this point are then measured using a combined sample of all exclusive decay modes. The distributions above 0.95 are parameterised using cubic spline functions $s_1(x_1)$ and $s_2(x_2)$, where $x_1$ and $x_2$ represent the BDT1 and BDT2 values. The combined efficiency for $x_1 > \alpha$ and $x_2 > \beta$ cuts is given by
\begin{align}
    \epsilon' &= \epsilon(x_1 > \alpha, x_2 > \beta \hspace{0.1cm}|\hspace{0.1cm} x_1 > 0.95, x_2 > 0.95) \nonumber \\
    &= \frac{\int_\alpha^{m_1}s_1 dx_1}{\int_{0.95}^{m_1} s_1 dx_1} \times \frac{\int_\beta^{m_2}s_2 dx_2}{\int_{0.95}^{m_2} s_2 dx_2},
\end{align}
where $m_1$ and $m_2$ are the maximum BDT1 and BDT2 scores observed in the summed background sample, respectively. The total efficiency for each background mode is then given by \mbox{$\epsilon = \epsilon(x_1 > 0.95, x_2 > 0.95) \times \epsilon'$}. To aid the spline descriptions of the remaining BDT distributions, the transformation $x \to -\log(1 - x)$ is performed to the BDT values $x$. The use of a combined sample of exclusive decays is justified by the observation that the BDT distributions beyond 0.95 are similar across all exclusive modes considered. 
\newpage
The spline fits to the summed exclusive background BDT distributions are shown in Fig.~\ref{fig:bkg_BDT_splines}, where an example cut of $> 0.99$ is shown along with the optimal cuts of BDT1 $> 0.99979$ and \mbox{BDT2 $> 0.99693$} found by the optimisation procedure. At these optimal cut values, the following yields are estimated for $N_Z = 5 \times 10^{12}$:
\begin{itemize}
    \item $N(B_c^+ \to \tau^+ \nu_\tau) = 4295$;
    \item $N(B^+ \to \tau^+ \nu_\tau) = 285$;
    \item Background = 448,
\end{itemize}
where the expected $\BTauNu$ yield is calculated using the measured branching fraction for this mode~\cite{PDG}. The total signal efficiency is found to be $0.39\%$, and the signal purity is determined to be 85\%, demonstrating the excellent performance of the two BDTs in reducing the background from $b$-hadron decays. The 6.6\% rate of the $\BTauNu$ mode relative to signal is also notable; given the factor $10^3$ higher production rate of $B^+$ mesons relative to $B_c^+$ mesons according to Pythia, and a relative branching ratio \mbox{$\mathcal{B}(B^+ \to \tau^+ \nu_\tau)/\mathcal{B}(B_c^+ \to \tau^+ \nu_\tau) \sim 0.5\%$}, the $B^+$ mode is expected to contribute at five times the level of the signal decay prior to any selection cuts. The full selection efficiency for the $\BTauNu$ mode is found to be $4.3 \times 10^{-5}$, owing to the excellent rejection of BDT2 in particular. The efficiencies for each exclusive background mode considered are given in App.~\ref{app:bkg_effs} Table~\ref{tab:bkg_BDT_effs}; efficiencies at the $10^{-10}-10^{-9}$ level are found.

\subsection{Fit to measure the signal yield} 

\begin{figure}[h!]
\centering
\includegraphics[width = 0.49\textwidth]{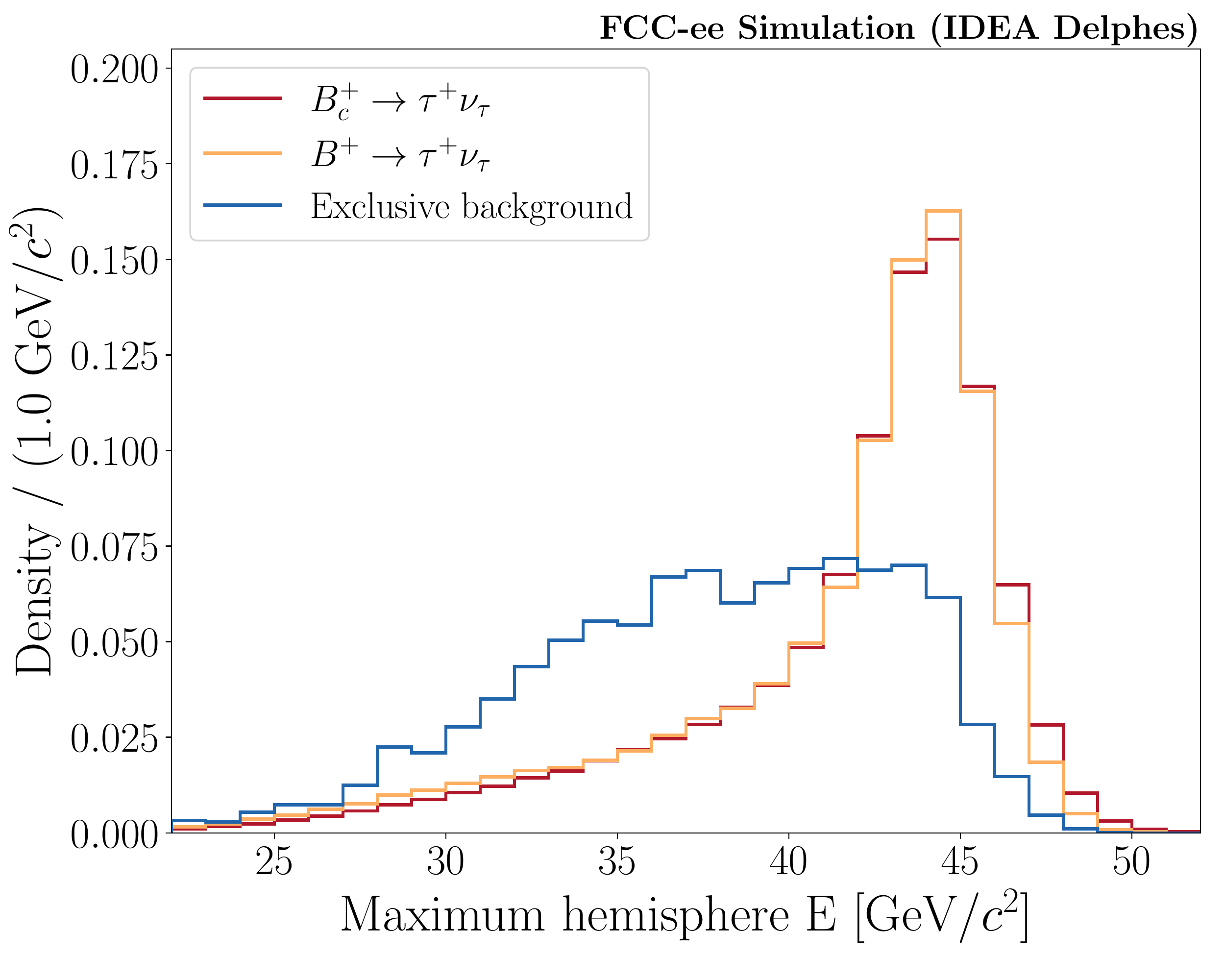} \put(-30,140){(a)}\\
\includegraphics[width = 0.42\textwidth]{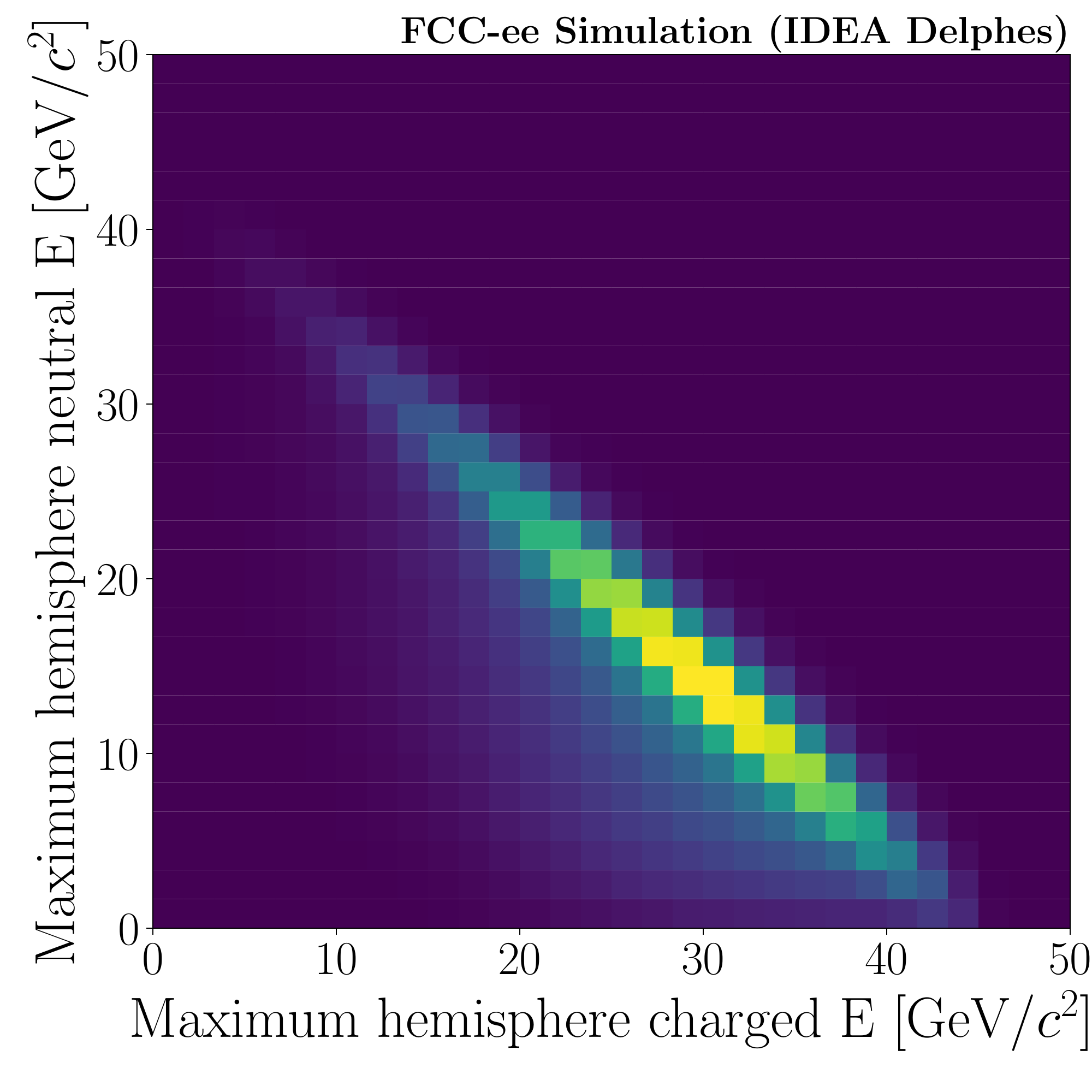}\put(-35,150){\textcolor{white}{(b)}}
\includegraphics[width = 0.42\textwidth]{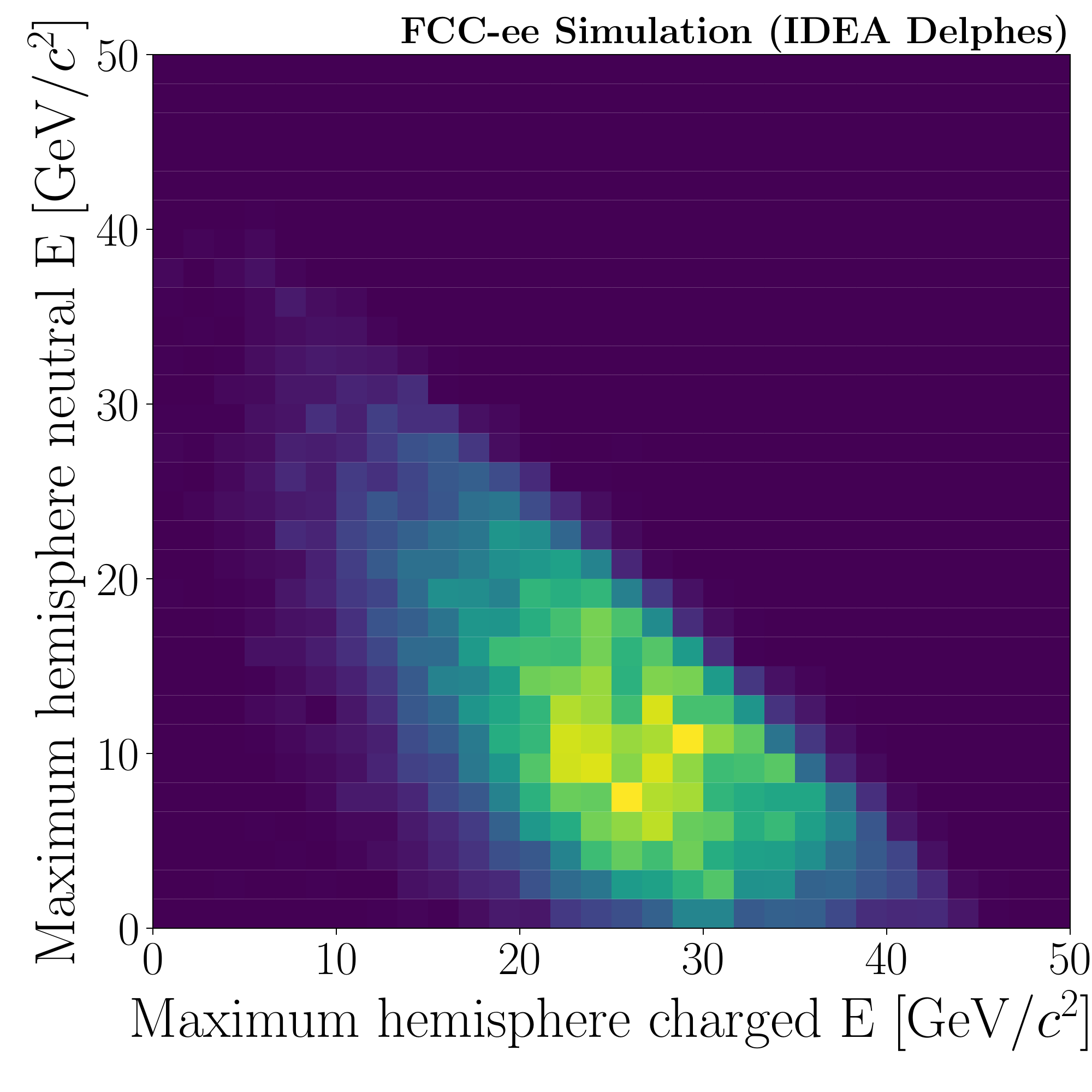}\put(-35,150){\textcolor{white}{(c)}}
\caption{(a) Distribution of total hemisphere energy for the maximum energy hemisphere. Signal and $\BTauNu$ decays closely follow the expected distribution for an inclusive $b$-quark decay from a $Z$, whereas the background distribution is biased downwards by the selection. (b/c) Relationship between the total charged and neutral energy in the maximum energy hemisphere for $\BcTauNu$/ inclusive $Z \to b\bar{b}$ events.}
\label{fig:Emax}
\end{figure}

To evaluate the potential precision of a signal yield measurement with $N_Z = 5 \times 10^{12}$, pseudoexperiment fit studies are performed. To select a fit variable, comparisons between all signal and background variable distributions are performed after tight BDT cuts; only those variables related to hemisphere energy are found to provide discrimination. Of all considered variables, the total energy in the maximum energy hemisphere is found to provide the most discrimination. Normalised distributions of this variable in signal events, $\BTauNu$ events, and exclusive background events are shown in Fig.~\ref{fig:Emax} (a), where cuts of $>0.99$ are applied to both BDTs. The exclusive background distribution shown is a sum of all exclusive $b$-hadron modes considered, where the modes are combined according to their expected yields from the cut optimisation.

In the selection, the signal $3\pi$ candidate is required to reside in the minimum energy hemisphere, since this is most likely to be the true signal hemisphere due to the large missing energy. As a result, in signal events the energy distribution in the maximum energy hemisphere closely resembles an inclusive $b$-quark decay from a $Z$ boson; a peak close to $m(Z)/2$ with a tail extending to lower energies due to missing energy. A very similar distribution is also observed in $\BTauNu$ events. The total charged and neutral energies in the maximum energy hemisphere are found to correspond closely in $B_{(c)}^+ \to \tau^+ \nu_\tau$ events, as shown in Fig.~\ref{fig:Emax} (b). The narrow diagonal observed corresponds to the peaking structure in  total hemisphere energy close to $m(Z)/2$, as shown in Fig.~\ref{fig:Emax} (a).

In $Z \to b\bar{b}$ background events, the distinction between the minimum and maximum energy hemispheres is not well defined, as either side of the event could contain more missing energy. Prior to any selection cuts, both hemispheres exhibit the inclusive structure with a peak close to $m(Z)/2$ and a tail. However, after application of the full selection, the energy distribution in the maximum energy hemisphere is biased downwards and the peaking structure is no longer observed. Both the charged and neutral energies in the maximum energy hemisphere are biased downwards by the selection, as shown in Fig.~\ref{fig:Emax} (c). The weaker correspondence between the neutral and charged energies along the diagonal results in no clear peaking structure and a spread to lower values, as shown in Fig.~\ref{fig:Emax} (a). These changes to the energy distributions are found both in the inclusive $Z \to b\bar{b}$ sample and in the summed exclusive $b$-hadron sample. For the pseudoexperiment fit studies, the exclusive sample is used in order to retain sufficient statistics for the creation of a template.

To create a total probability density function (PDF) for the pseudoexperiment fit, the normalised histogram templates shown in Fig.~\ref{fig:Emax} are each multiplied by the corresponding yield estimates from the cut optimisation and then summed. To create a pseudoexperiment dataset, a copy of the total PDF is created and each bin varied independently according to Poisson statistics. A fit is then performed to the pseudoexperiment dataset using the total PDF, where the yield of the signal and background components are free parameters. As the $\BTauNu$ distribution is very similar to signal, and this mode contributes at only the $\sim 7\%$ level relative to signal, it is not possible to freely vary $N(B^+ \to \tau^+ \nu_\tau)$ in the fit. As such, the yield of this component is constrained according to a Gaussian, with a central value of 249 events (the expected yield from the optimisation) and a width of 10 events. This width corresponds to 5\% relative uncertainty on the $\BTauNu$ yield, which is the anticipated precision on $\mathcal{B}(B^+ \to \tau^+ \nu_\tau)$ from Belle II~\cite{BelleII}. Given the small anticipated contribution from the $\BTauNu$ mode relative to signal, the absolute uncertainty on $\mathcal{B}(B^+ \to \tau^+ \nu_\tau)$ will not contribute a significant source of uncertainty to $N(B_c^+ \to \tau^+ \nu_\tau)$.

\begin{figure}[h!]
\centering
\includegraphics[width = 0.49\textwidth]{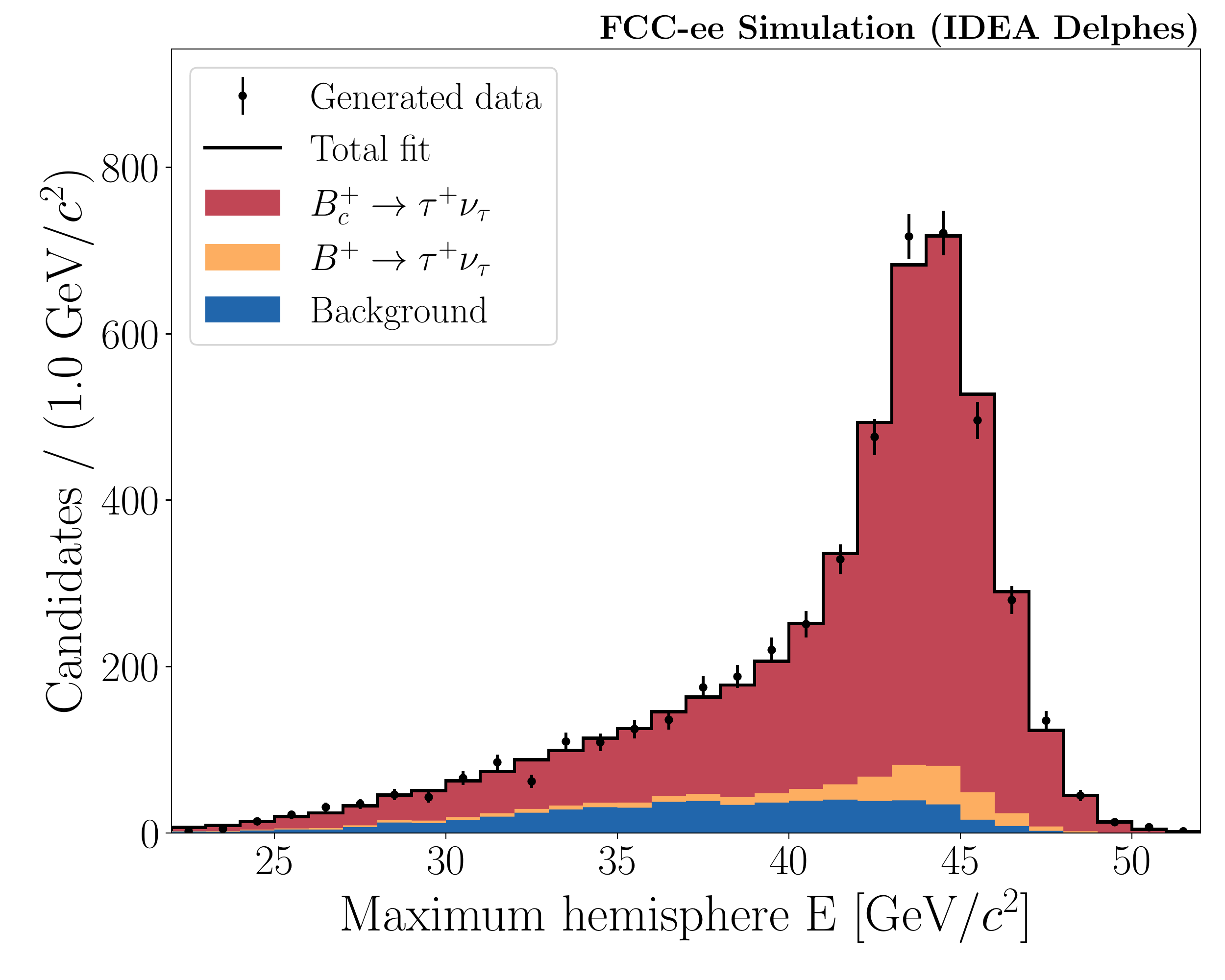}\put(-30,140){(a)}
\includegraphics[width = 0.39\textwidth]{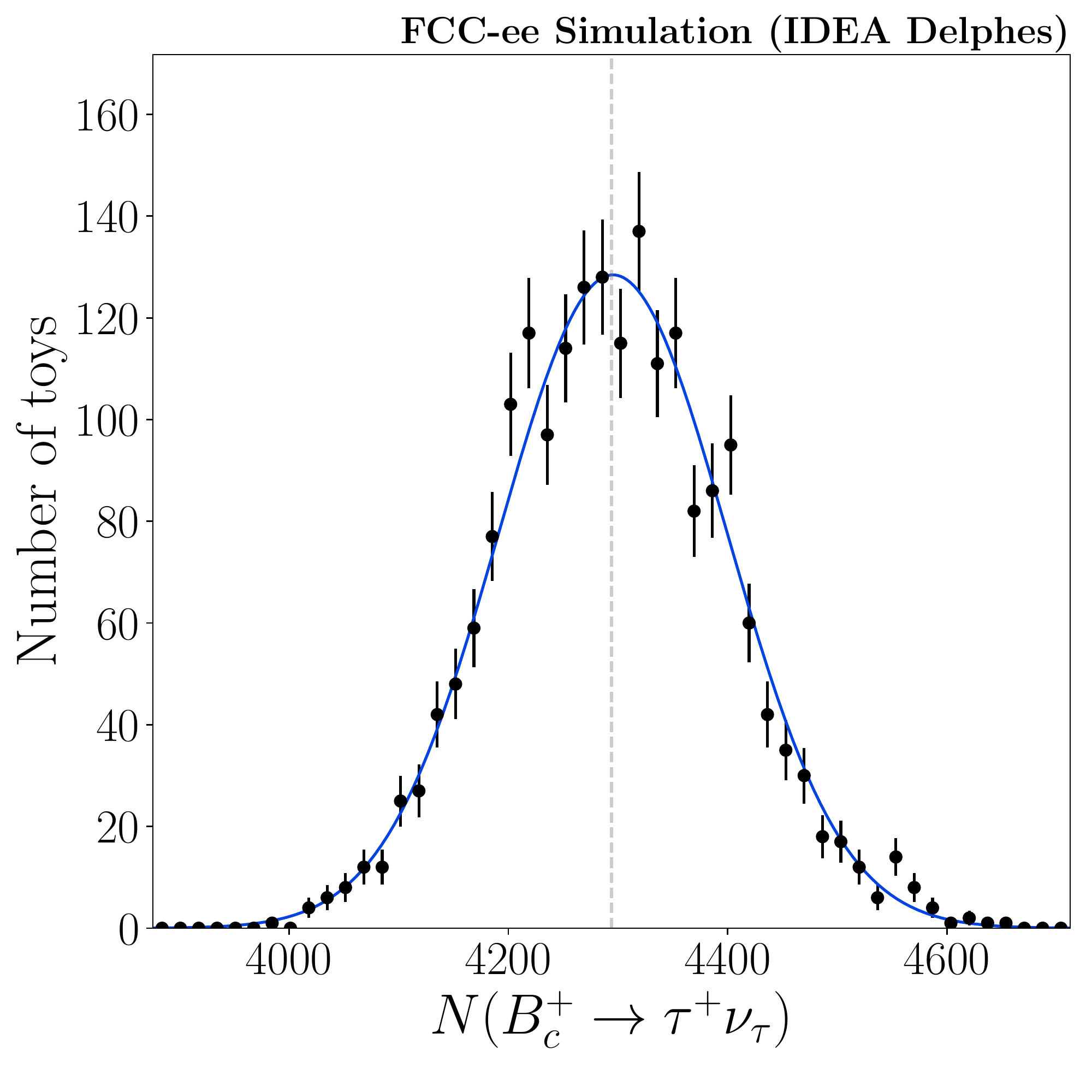}\put(-30,140){(b)}
\caption{(a) Result of a single pseudoexperiment fit, where the peaking signal is clearly distinguishable from the background. (b) Signal yields measured in 2000 pseudoexperiment fits, where the generated value is indicated by the dashed vertical line.}
\label{fig:fit}
\end{figure}

The result of a single pseudoexperiment fit is shown in Fig.~\ref{fig:fit} (a), and the distribution of signal yields measured in 2000 fits to different generated pseudoexperiment datasets is shown in Fig.~\ref{fig:fit} (b). Both the signal and background yields are measured without bias in either their central values or uncertainties. From a fit to the distribution of signal yields, the signal yield uncertainty is found to be 101, which corresponds to a relative uncertainty of 2.4\% on the generated yield of 4295 events. This uncertainty is purely statistical, and does not account for potential sources of systematic uncertainty. However, the expected 5\% relative uncertainty on $\BFBTauNu$ is included via the Gaussian constraint on $N(B^+ \to \tau^+ \nu_\tau)$ in the fits. Fits performed with ten times more background included in the toy samples are also found to be stable, with a relative signal yield uncertainty of 2.9\%. The analysis is thus robust to large increases in background yield relative to what is modelled here.

\subsection{Fit performance for different $N_Z$}

The fit results shown in Fig.~\ref{fig:fit} correspond to a final sample selected from a dataset containing $N_Z = 5 \times 10^{12}$. The fit performance is also studied at lower sample sizes of $N_Z = [0.5, 1, 2, 3, 4] \times 10^{12}$, in order to evaluate the signal yield precision possible at earlier stages of FCC-ee operation. For each $N_Z$ value, the cut optimisation is rerun in order to maximise the purity; highly consistent optimal purity is found across all $N_Z$ values. Sets of 2000 pseudoexperiment fits are run for each $N_Z$ value, using the expected signal, $\BTauNu$, and background yields from the cut optimisation. 

\begin{figure}[h!]
\centering
\includegraphics[width = 0.49\textwidth]{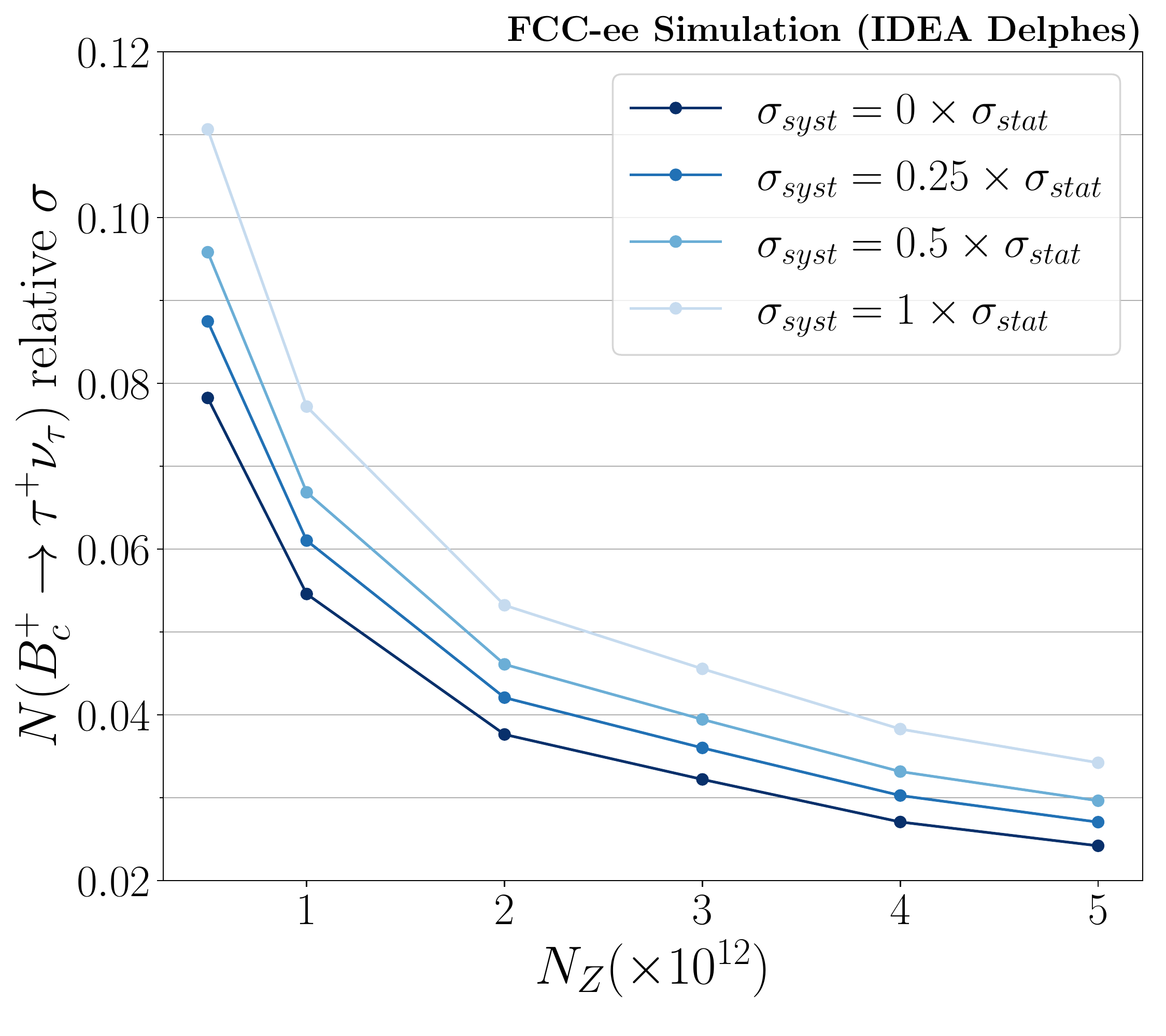}
\caption{Relative precision on the signal yield as a function of $N_Z$. The signal yields at each $N_Z$ value are taken from the cut optimisation procedure, and the statistical uncertainties are measured in pseudoexperiment fits. Different levels of systematic uncertainty relative to the statistical uncertainty are also shown.} 
\label{fig:NBc_vs_NZ}
\end{figure}

The signal yields expected as a function of $N_Z$, as well as their uncertainties as measured in the pseudoexperiment fits, are summarised in Tab.~\ref{tab:NBc_vs_NZ}. The relative signal yield precision as a function of $N_Z$ is illustrated in Fig.~\ref{fig:NBc_vs_NZ}, where four different systematic uncertainty scenarios are shown; $\sigma_{\text{syst}} = [0, 0.25, 0.5, 1] \times \sigma_{\text{stat}}$. All values shown are summarised in App.~\ref{app:precisions} Tab.~\ref{tab:N_Bc2TauNu_vs_NZ}. The level of systematic uncertainty in a real analysis will depend on several factors, such as: 
\begin{itemize}
    \item Detector resolution, reconstruction efficiency, and calibration quality;
    \item  The size of the simulated samples used to create fit templates;
    \item The decay models used to generate signal and background decays;
    \item Knowledge of the relative proportions of decay modes entering the total background template.
\end{itemize}
Given the high signal purity achievable, however, and the distinctive shape of the signal maximum hemisphere energy distribution, an eventual measurement is not expected to be limited by systematic uncertainties. Assuming that the systematic uncertainties can be controlled at the level $\sigma_{\text{syst}} = \sigma_{\text{stat}}$, the  relative precision possible on $N(B_c^+ \to \tau^+ \nu_\tau)$ with $N_Z = 5 \times 10^{12}$ is $\sqrt{2 \times 104^2}/4295 = 3.4\%$. 

\renewcommand{\arraystretch}{1.2}{
\begin{table}[]
\centering
\begin{tabular}{ccc}
$N_Z (\times 10^{12})$ & $N(B_c^+ \to \tau^+ \nu_\tau)$ & Relative $\sigma$ (\%) \\ \hline
0.5 & $430\phantom{0} \pm 33\phantom{0}$ & 7.8 \\
1 & $858\phantom{0} \pm 46\phantom{0}$ & 5.5 \\
2 & $1717 \pm 64\phantom{0}$ & 3.8 \\
3 & $2578 \pm 83\phantom{0}$ & 3.2 \\
4 & $3436 \pm 93\phantom{0}$ & 2.7 \\
5 & $4295 \pm 103$ & 2.4 \\
\hline
\end{tabular}
\caption{Estimated signal yields as a function of $N_Z$, where the uncertainties quoted are statistical only. The yield central values are determined from the cut optimisation procedure, and the uncertainties from pseudoexperiment fits.}
\label{tab:NBc_vs_NZ}
\end{table}
}

\subsection{Branching fraction determination}

It is common to measure signal modes relative to a normalisation decay, in order to minimise systematic uncertainties and cancel the effects of hadron production. One suitable choice of normalisation mode for $\BcTauNu$ is the semileptonic $B_c^+ \to J/\psi \mu^+ \nu_\mu$ decay, where the \mbox{$J/\psi \to \mu^+ \mu^-$} channel can be selected in order to provide a clean three-muon $B_c^+$ decay vertex. This mode can be reconstructed and selected with high efficiency, as sources of lighter $b$-hadron background can be eliminated with a $m(J/\psi \mu) > 5.3$ GeV/$c^2$ cut~\cite{Bc2JpsiMuNu_Bc2JpsiPi_LHCb,JpsiTauNu_LHCb}. Above this cut, the only sources of remaining background are from random combinations of three muons (expected to be small at a FCC-ee) and contributions from $B_c^+ \to J/\psi \mu^+ \nu_\mu X$ decays where $X$ is not considered in the invariant mass sum. The latter contribution, from decays such as $B_c^+ \to (\psi(2S) \to J/\psi \pi^+ \pi^-) \mu^+ \nu_\mu$, can be reduced using isolation requirement, where all other charged particles and neutrals in the signal hemisphere must be inconsistent with originating from the $3\mu$ vertex. 

With $B_c^+ \to J/\psi \mu^+ \nu_\mu$ as a normalisation mode, the ratio of branching fractions
\begin{align}
R_c &= \frac{\mathcal{B}(B_c^+ \to \tau^+ \nu_\tau)}{\mathcal{B}(B_c^+ \to J/\psi \mu^+ \nu_\mu)} \nonumber \\
&= \frac{N(B_c^+ \to \tau^+ \nu_\tau)}{N(B_c^+ \to J/\psi \mu^+ \nu_\mu)} \times \frac{\epsilon(B_c^+ \to J/\psi \mu^+ \nu_\mu)}{\epsilon(B_c^+ \to \tau^+ \nu_\tau)} \times \frac{\mathcal{B}(J/\psi \to \mu^+ \mu^-)}{\mathcal{B}(\TauThreePi)}
\end{align}
can be measured. Assuming SM amplitudes in the normalisation decay, the ratio $R_c$ is highly sensitive to NP couplings to $\tau$ leptons, and is independent of the value of $|V_{cb}|$ and the $B_c^+$ hadronisation fraction $f(B_c^+)$. The $J/\psi$ and $\tau$ branching fractions are well measured~\cite{PDG}, while the signal and normalisation efficiencies can be determined with high accuracy given sufficiently large simulated samples. As the signal and normalisation modes both involve the reconstruction and selection of three charged tracks from a common vertex, systematic uncertainties in the absolute efficiencies are expected to cancel to a high degree in the ratio. The SM prediction for \mbox{$\mathcal{B}(B_c^+ \to J/\psi \mu^+ \nu_\mu)^\mathrm{SM} = 0.0135 \pm 0.0011$}, which is calculated using lattice QCD and the current  value of $|V_{cb}|^{\mathrm{excl.}}$~\cite{Aoki:2019cca}. Assuming a selection efficiency of 10\% for this mode, the anticipated normalisation yield is around $50,000$ events with $N_Z = 5 \times 10^{12}$. The normalisation yield is thus an order of magnitude larger than the signal yield, and as such will not contribute significantly to the branching fraction ratio uncertainty. Thus, the precision on $R_c$ is expected to be dominated by the uncertainty on $N(B_c^+ \to \tau^+ \nu_\tau)$. 

\begin{figure}[h!]
\centering
\includegraphics[width = 0.45\textwidth]{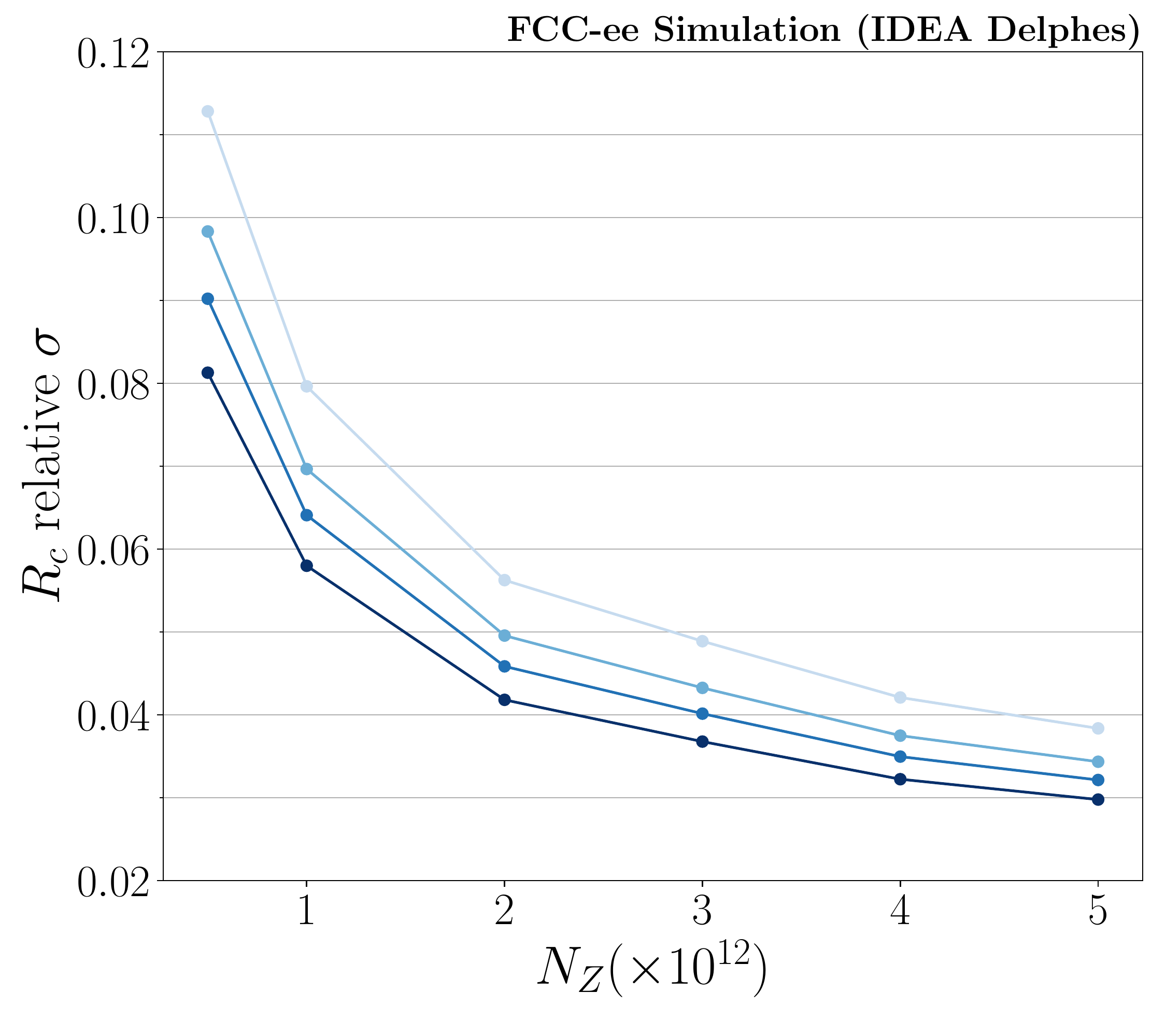}\put(-30,140){(a)}
\includegraphics[width = 0.45\textwidth]{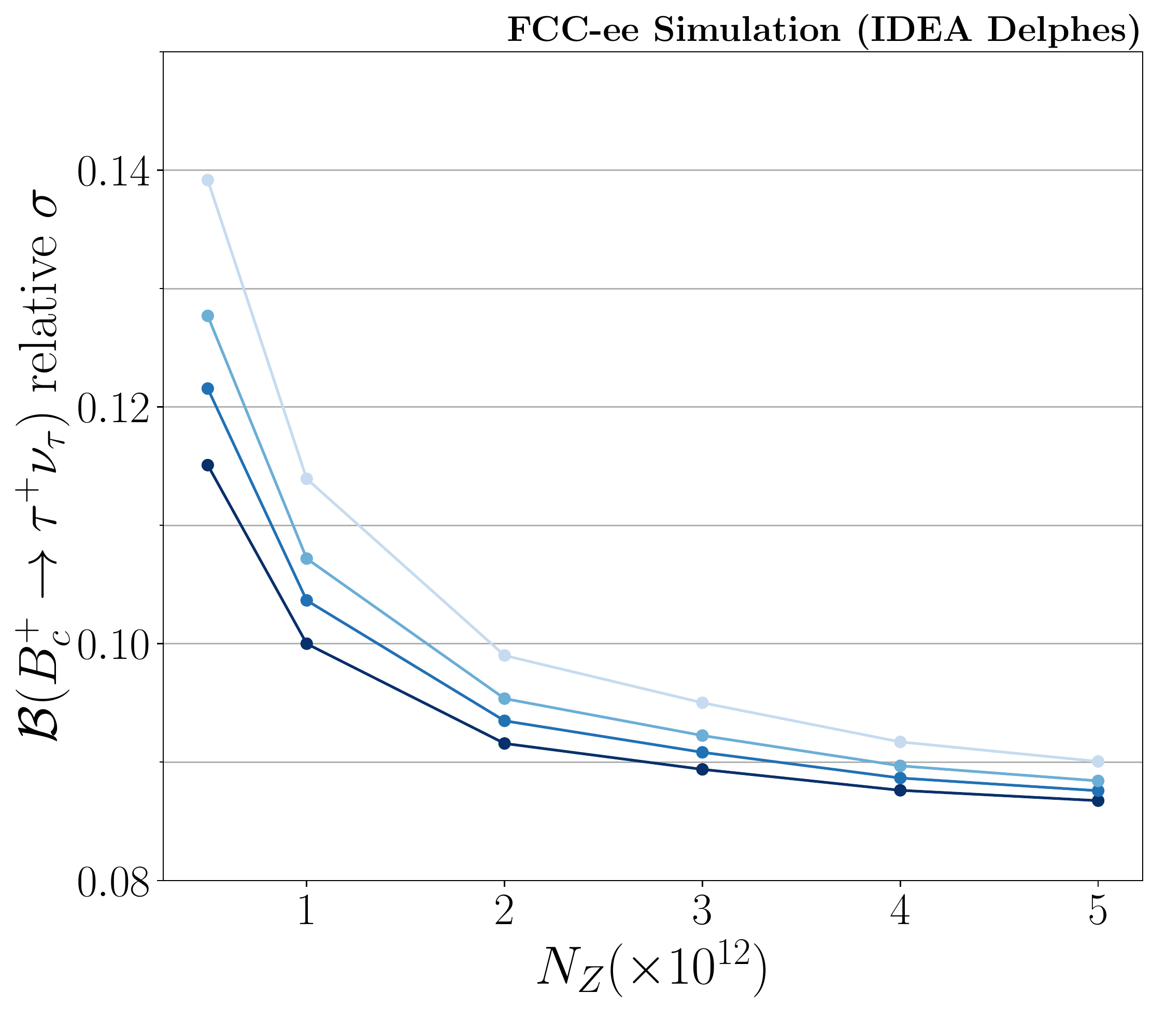}\put(-30,140){(b)}
\caption{(a) Relative precision on the ratio of branching fractions \mbox{$R_c = \mathcal{B}(B_c^+ \to \tau^+ \nu_\tau)/\mathcal{B}(B_c^+ \to J/\psi \mu^+ \nu_\mu)$} as a function of $N_Z$. (b) Relative precision on $\BFBcTauNu$ as a function of $N_Z$, using a SM prediction for $\mathcal{B}(B_c^+ \to J/\psi \mu^+ \nu_\mu)$. The different shades of blue correspond to different levels of systematic uncertainty on $N(B_c^+ \to \tau^+ \nu_\tau)$ relative to the statistical uncertainty, following the same colour scheme as Fig.~\ref{fig:NBc_vs_NZ}.}
\label{fig:BF_vs_NZ}
\end{figure}

The anticipated relative uncertainty on $R_c$ as a function of $N_Z$ is shown in Fig.~\ref{fig:BF_vs_NZ} (a), where the signal yields and uncertainties from Tab.~\ref{tab:NBc_vs_NZ} are used as input. The uncertainties shown also include the current uncertainties on the $J/\psi$ and $\tau$ branching fractions, a $1\%$ relative uncertainty on both the signal and normalisation mode efficiencies, and a $\sqrt{N}$ uncertainty on the normalisation yield. The relative precision on $R_c$ is found to closely follow the relative precision on $N(B_c^+ \to \tau^+ \nu_\tau)$. All values shown are summarised in App.~\ref{app:precisions} Tab.~\ref{tab:BF_ratio_vs_NZ}.

\sloppypar
It is also possible to determine an absolute branching fraction for the signal decay,
\begin{equation}
    \mathcal{B}(B_c^+ \to \tau^+ \nu_\tau) = R_c \times \mathcal{B}(B_c^+ \to J/\psi \mu^+ \nu_\mu)^{\mathrm{SM}},
\end{equation}
where the measured ratio of branching fractions $R_c$ is multiplied by the SM prediction for \mbox{$\mathcal{B}(B_c^+ \to J/\psi \mu^+ \nu_\mu)$}. In doing so, one must assume a value of $|V_{cb}|$ in the calculation of the normalisation branching fraction; in such a setup, an interpretation of $\mathcal{B}(B_c^+ \to \tau^+ \nu_\tau)$ in terms of $|V_{cb}|$ would not be plausible. Naturally, the precision on the absolute branching fraction $\BFBcTauNu$ will be impacted by the precision on the normalisation mode branching fraction prediction, which at present is 8\% relative. The resulting limitation in \mbox{$\BFBcTauNu$} prediction can be seen in Fig.~\ref{fig:BF_vs_NZ} (b), where the improvement with $N_Z$ is more modest than for $R_c$ due to the additional uncertainty from $\mathcal{B}(B_c^+ \to J/\psi \mu^+ \nu_\mu)$; all values shown are summarised in App.~\ref{app:precisions} Tab.~\ref{tab:BF_Bc2TauNu_vs_NZ}. The precision of the normalisation mode branching fraction calculation is limited at present by knowledge of the $B_c^+ \to J/\psi$ form factors, which can be improved in future measurements of the $B_c^+ \to J/\psi \mu^+ \nu_\mu$ decay both at LHCb and FCC-ee. In particular, complete form factor information could be determined from an angular analysis, following the approach described in Ref.~\cite{Hill:2019zja} to measure a full set of angular coefficients. Assuming SM amplitudes in the decay would enable form factor information to derived directly from the measured angular coefficients.


\subsection{Additional factors to consider}

In the analysis presented above, a parametric description of the IDEA detector is employed to model key elements of the expected detector response such as momentum and impact parameter resolution. In future, studies using full simulation will be required to evaluate the complete resolution on key quantities such as the hemisphere energies, as well as to determine the reconstruction and selection efficiencies expected in genuine FCC-ee operation. Such studies should be performed under different detector design scenarios, in order to determine how aspects such as tracking, calorimetry, and vertex reconstruction may influence the expected precision on $N(B_c^+ \to \tau^+ \nu_\tau)$.

The impact of particle identification (PID) should also be studied, since the current analysis is performed using combinations of three genuine charged pions only i.e.~perfect PID is assumed. The pions produced in $\BcTauNu$ decays have momenta in the $1-10$ GeV/$c$ range, where $dE/dx$, time of flight, and Cherenkov techniques can provide high discrimination between pions, kaons, and muons. Muon rejection is necessary to suppress high branching fraction semileptonic decays of beauty and charm hadrons, while kaon rejection is important since kaons are often produced in the decays of charm hadrons to multi-track final states.

It may be possible to extend the measurement to include the $\tau^+ \to \pi^+ \pi^+ \pi^- \pi^0 \bar{\nu}_\tau$ mode, which also provides a $3\pi$ vertex and has a branching fraction of 50\% relative to the \mbox{$\tau^+ \to \pi^+ \pi^+ \pi^- \bar{\nu}_\tau$} decay. Due to differences in $3\pi$ kinematics and resonant structure, the selection employed in this analysis is highly inefficient on the $\tau^+ \to \pi^+ \pi^+ \pi^- \pi^0 \bar{\nu}_\tau$ decay, rendering any contribution in the final sample negligible. As such, a dedicated selection using \mbox{$B_c^+ \to \tau^+ \nu_\tau$} with \mbox{$\tau^+ \to \pi^+ \pi^+ \pi^- \pi^0 \bar{\nu}_\tau$} signal MC for the MVA training and cut optimisation would be required in order to isolate these decays. Such a strategy could either include reconstruction of the additional neutral pion, or proceed with only $3\pi$ vertex reconstruction. In either case, the contribution from $3\pi$ signal decays passing the $3\pi \pi^0$ selection must be modelled.

In the analysis presented within, only combinations of pions originating from common decay vertices are considered. In real data, combinatorial background will also contribute, where one of more of the pions in the reconstructed $3\pi$ system will not originate from a common vertex. Requirements on vertex $\chi^2$, charged track impact parameter from the primary vertex, and track momentum, should assist in minimising such contributions. The remaining level of combinatorial background in data can be estimated using combinations of particles such $\pi^+ \pi^+ \pi^+$, which are non-physical and thus represent purely random combinatorics.

Finally, samples of both inclusive and exclusive background decays should be generated at the levels expected in a dataset of size $N_Z \sim 10^{12}$. This will enable background rejections to be accurately measured up to the required $10^{10}$ level, with sufficient statistics remaining to model the background contributions in the signal yield fit.

\FloatBarrier

\section{Implications for New Physics}
\label{sec:interpretation}

The phenomenological impact of a measurement of $\mathcal{B}(B_c^+\to \tau^+ \nu_\tau)$ with the precision depicted in Fig.~\ref{fig:BF_vs_NZ} is now explored for a few NP scenarios. A measurement of the ratio \mbox{$R_c = \mathcal{B}(B_c^+\to\tau^+ \nu_\tau)/ \mathcal{B}(B_c^+ \to J/\psi \mu^+\nu_\mu)$} with a precision of $\approx 4\%$ is considered, and it is assumed that $\mathcal{B}(B_c^+\to J/\psi \mu^+ \nu_\tau)$ is not affected by NP contributions, which is a well justified assumption for the models discussed below. For the external input $\Gamma(B_c^+ \to J/\psi \mu^+\nu_\mu)/|V_{cb}|^2$, we consider two benchmark scenarios: (i) a relative uncertainty of $\approx 7 \%$, as currently obtained with LQCD form factors, and (ii) an uncertainty of $\approx 2 \%$ which could be obtained in the future by combining LQCD and experimental inputs, as discussed in Sec.~\ref{sec:analysis}. These two scenarios amount to an uncertainty on $\Gamma(B_c^+\to\tau^+ \nu_\tau)/|V_{cb}|^2$ of $\approx 10 \%$ and $\approx 4 \%$, respectively, which will be considered in what follows to constrain NP contributions.~\footnote{Note that the partial decay width $\Gamma(B_c^+ \to \tau \nu_\tau)$ is used instead of $\mathcal{B}(B_c^+\to \tau^+ \nu_\tau)$, since the ratio $R_c$ is independent of $\tau_{B_c}$ which would imply an additional source of uncertainty ($\approx 2\%$)~\cite{Zyla:2020zbs} to the interpretation of $R_c$ if the latter was considered.}

\subsection{Effective Hamiltonian}

Firstly, the most general dimension-six effective Hamiltonian encoding SM and NP contributions to the $b\to c \tau \nu_\tau$ transition is considered~\cite{Becirevic:2020rzi},
\begin{align}
\label{eq:left}
    \mathcal{H}_\mathrm{eff} &= 2\sqrt{2}G_F V_{cb}\Big{[}(1+g_{V_L})\,\big{(}\bar{c}_{L}\gamma_\mu b_{L} \big{)}\big{(}\bar{\tau}_L \gamma^\mu\nu_{L}\big{)}+g_{V_R}\,\big{(}\bar{c}_{R}\gamma_\mu b_{R} \big{)}\big{(}\bar{\tau}_L \gamma^\mu\nu_{L}\big{)}\\[0.38em]
    &+g_{S_L}\,\big{(}\bar{c}_{R} b_{L} \big{)}\big{(}\bar{\tau}_R \nu_{L}\big{)}+g_{S_R}\,\big{(}\bar{c}_{L} b_{R} \big{)}\big{(}\bar{\tau}_R\nu_{L}\big{)}+g_{T}\,\big{(}\bar{c}_{R}\sigma_{\mu\nu} b_{L} \big{)}\big{(}\bar{\tau}_R \sigma^{\mu\nu}\nu_{L}\big{)}\,\Big{]}+\mathrm{h.c.}\,,\nonumber
\end{align}

\noindent where $g_{\alpha}\equiv g_{\alpha}(\mu)$ with $\alpha\in \lbrace V_{L(R)},S_{L(R)},T\rbrace$ represent the effective coefficients evaluated at the renormalisation scale $\mu$, which is taken to be $\mu=m_b$ unless stated otherwise. The normalisation is such that the SM corresponds to $g_{\alpha}=0$ for all coefficients. The $B_c^+ \to \tau^+ \nu_\tau$ decay branching fraction can then be written in full generality in terms of Eq.~\eqref{eq:left},
\begin{align}
\label{eq:Bctaunu}
\mathcal{B}(B_c^+ \to \tau^+ \nu_\tau)= \mathcal{B}(B_c^+ \to \tau^+ \nu_\tau)^\mathrm{SM}\times\left|1-g_{A}+g_{P}\dfrac{m_{B_c}^2}{m_\tau(m_b+m_c)}\right|^2\,,
\end{align}

\noindent where $g_{A(V)} \equiv g_{V_R}\mp g_{V_L}$ and $g_{P(S)} \equiv g_{S_R}\mp g_{S_L}$ are defined. From Eq.~\eqref{eq:Bctaunu}, it is clear that the $B_{c}^+ \to \tau^+ \nu_\tau$ decays are particularly sensitive to pseudoscalar couplings of NP since they lift the helicity suppression of the SM amplitude. By considering an experimental precision of $\approx 10\%$ on $\Gamma(B_c^+\to\tau^+ \nu_\tau)/|V_{cb}|^2$ and assuming a central value that coincides with the SM prediction, the following sensitivity ($95\%~\mathrm{CL.}$) on the effective couplings is expected,
\begin{align}
    \label{eq:limit-10p}
    g_A \in (-0.10,0.11)\,,\qquad\qquad\quad g_P \in (-0.024,-0.024)\,,
\end{align}
where the couplings are assumed to be real. With an improved determination of the normalisation channel $B_c^+\to J/\psi \mu^+ \nu_\mu$, a precision of $\approx 4\%$ on $\Gamma(B_c^+\to\tau^+ \nu_\tau)/|V_{cb}|^2$ can be assumed, which would amount to
\begin{align}
    \label{eq:limit-4p}
    g_A \in (-0.05,0.05)\,,\qquad\qquad\quad g_P \in (-0.011,-0.011)\,.
\end{align}
Note, in particular, that such a measurement would considerably improve the sensitivity on the coupling $g_P$, which is only weakly constrained at present by the requirement that $\Gamma(B_c^+\to \tau^+ \nu_\tau)$ should not saturate $\approx 30 \%$ of the total $B_c^+$ meson width~\cite{Alonso:2016oyd,Li:2016vvp} which is determined experimentally~\cite{Zyla:2020zbs}.

The effective couplings defined in Eq.~\eqref{eq:left} can arise in several extensions of the SM. Of particular interest are extensions of the SM Higgs sector such as the Two-Higgs-doublet model (2HDM)~\cite{Branco:2011iw} and specific models containing scalar and vector leptoquarks (LQs)~\cite{Buchmuller:1986zs,Dorsner:2016wpm}, since they can induce the coefficients $g_P$ at low energies. The implications of the limits derived in Eq.~\eqref{eq:limit-10p} and \eqref{eq:limit-4p} for these scenarios are now discussed.

\subsection{2HDM}

One of the minimal extensions of the SM consists in enlarging the Higgs sector with an additional Higgs doublet with the same quantum numbers~\cite{Branco:2011iw}. Besides the SM-like Higgs boson, the spectrum of these models contain an extra $C\!P$-even Higgs, a neutral $C\!P$-odd scalar, as well as a charged Higgs boson that can contribute to charged-current transitions such as the one studied in this work. In order to avoid Flavor Changing Neutral Currents (FCNCs) from appearing at tree-level, one imposes via a discrete symmetry that fermions of a specific chirality and hypercharge should couple to a single Higgs doublet~\cite{Glashow:1976nt}. Four choices are then possible, which are known as 2HDM of type-I, II, X, and Y~\cite{Branco:2011iw}. Among those, the type-II 2HDM is a popular choice which is embedded in the Minimal Supersymmetric extension
of the Standard Model (MSSM), and which has a rich phenomenology in flavour-physics observables~\cite{Hou:1992sy,Misiak:2017bgg,Crivellin:2013wna,Li:2014fea,Arnan:2017lxi,Eberhardt:2020dat}.

\begin{figure}[!t]
\centering
\includegraphics[width=.6\linewidth]{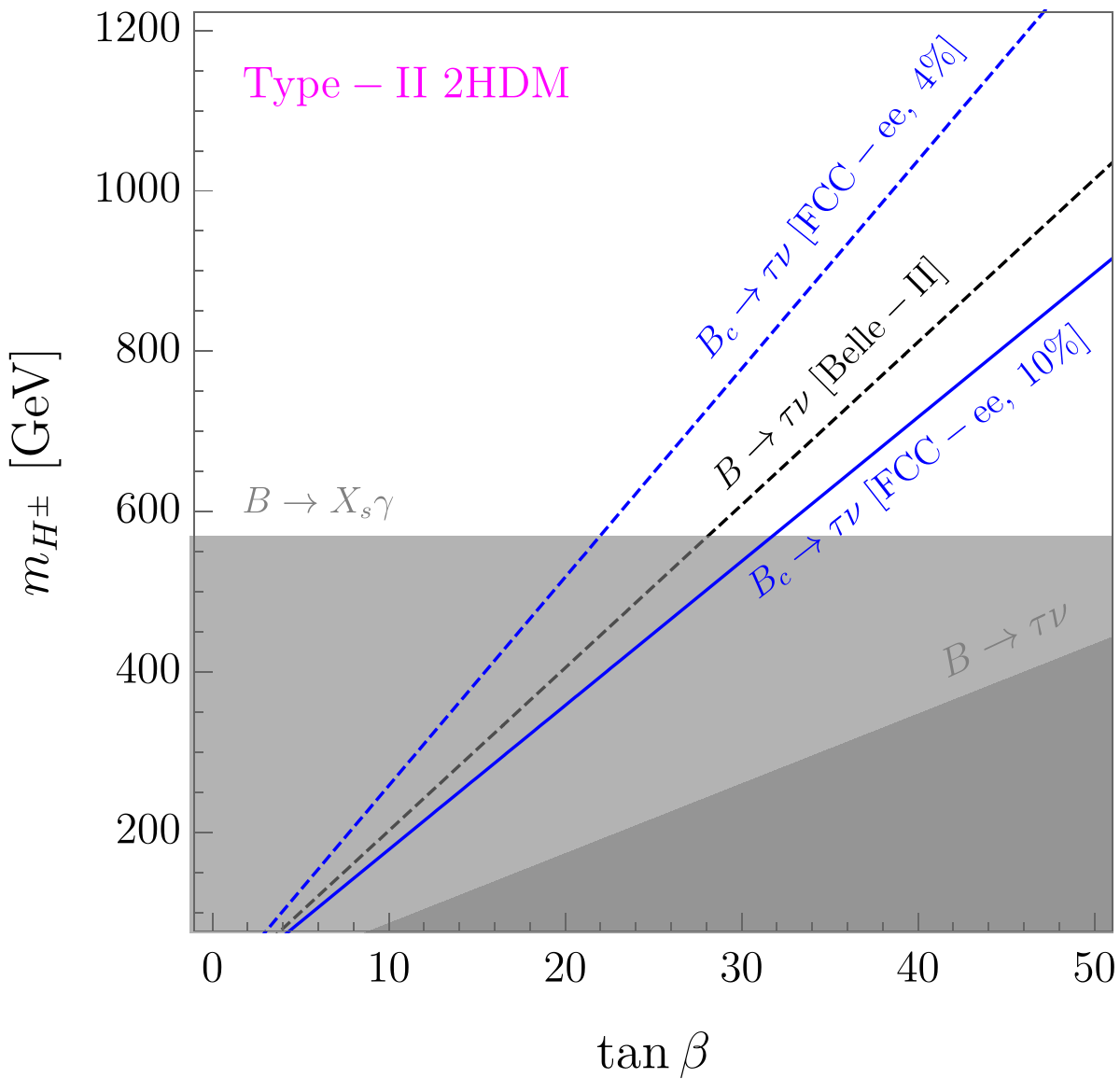}
\caption{Expected constraints on the plane $\tan \beta$ vs.~$m_{H^\pm}$ for type-II 2HDM derived by assuming a relative uncertainty on $\Gamma(B_c^+\to \tau^+ \nu_\tau)/|V_{cb}|^2$ of $10\%$ (solid blue line) and $4\%$ (dashed blue line). Current constraints obtained from $\mathcal{B}(B\to X_s \gamma)$~\cite{Misiak:2017bgg} and $\mathcal{B}(B^+\to \tau^+ \nu_\tau)$~\cite{Zyla:2020zbs} are depicted by the grey regions. Prospects for a $\mathcal{B}(B^+\to \tau^+ \nu_\tau)$ measurement at Belle-II are depicted by the grey dashed line, obtained under the assumption of $5\%$ uncertainty on the branching fraction~\cite{Kou:2018nap}.}
\label{fig:Bctaunu-2HDM} 
\end{figure}

The tree-level contribution of the charged Higgs ($H^\pm$) to the transition $b\to c\tau \nu_\tau$ can be matched to the effective Hamiltonian~\eqref{eq:left} generating the effective coefficients $g_S$ and $g_P$. For the type-II 2HDM, it is found that~\cite{Becirevic:2015fmu}
\begin{equation}
    g_P^{\mathrm{(II)}}= - \dfrac{m_\tau (m_{c}- m_b \tan^2 \beta)}{m_{H^\pm}^2}\,,~\qquad\quad g_S^{\mathrm{(II)}}= - \dfrac{m_\tau (m_{c} + m_b \tan^2 \beta)}{m_{H^\pm}^2}\,,
\end{equation}
\noindent where $m_{H^\pm}$ is the $H^\pm$ mass and $\tan\beta=v_2/v_1$ represents the ratio of the Higgs vacuum expectation values $v_{1,2}$, which satisfy $v^\mathrm{SM}=\sqrt{v_1^2+v_2^2}=246.2~\mathrm{GeV}$. The same expressions hold for example for the $b\to u \tau \nu_\tau$ transition by suitably replacing the quark masses.

The expected FCC-ee constraints on the plane $\tan\beta$ vs.~$m_{H^\pm}$ are shown in Fig.~\ref{fig:Bctaunu-2HDM} by using the limits on $g_P$ derived in Eq.~\eqref{eq:limit-10p} and \eqref{eq:limit-4p} with the assumption of $10\%$ and $4\%$ experimental sensitivity on the value of $\Gamma(B_c^+\to\tau^+\nu_\tau)/|V_{cb}|^2$, respectively. In the same plot, the constraints derived from $\mathcal{B}(B\to X_s \gamma)$~\cite{Misiak:2017bgg} are superimposed, which set a lower limit $m_{H^\pm} \gtrsim 570~\mathrm{GeV}$ at $95\%$~CL. Furthermore, the constraints arising from the current measurement of $\mathcal{B}(B^+\to \tau^+ \nu_\tau)$~\cite{Zyla:2020zbs} are shown, as well as the future prospects at Belle-II where a precision of $5\%$~\cite{Kou:2018nap} is anticipated. From this plot, it is clear that the measurement of $B_c^+\to\tau^+\nu_\tau$ at FCC-ee can probe an important region of parameter space in type-II 2HDM which will not be covered by other flavour constraints.

\subsection{Leptoquarks}

\begin{figure}[!t]
\centering
\includegraphics[width=.5\linewidth]{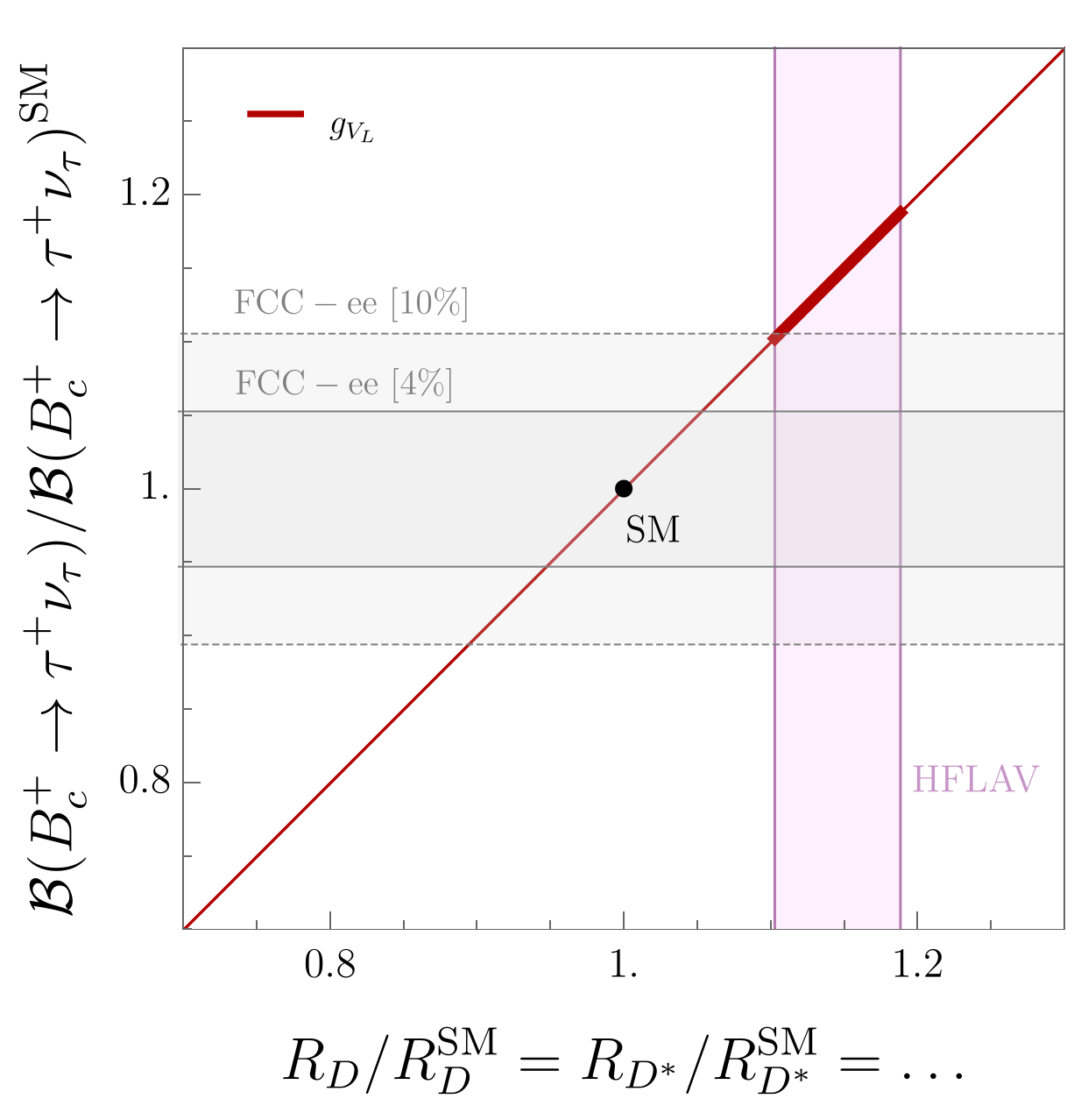}
\caption{Predictions for $R_{D^{(\ast)}}/R_{D^{(\ast)}}^\mathrm{SM}$ are plotted against the ratio \mbox{$\mathcal{B}(B_c^+\to\tau^+ \nu_\tau)/\mathcal{B}(B_c^+\to\tau^+ \nu_\tau)^\mathrm{SM}$} in the scenario with nonzero values of $g_{V_L}$. The thick lines correspond to the values of the effective couplings favoured by the current fit to $b\to c\tau \nu_\tau$ data~\cite{Angelescu:2021lln}. The magenta shaded region denotes the current average of $R_D^\mathrm{exp}/R_{D}^\mathrm{SM}$ and $R_{D^\ast}^\mathrm{exp}/R_{D^\ast}^\mathrm{SM}$ at $1\sigma$ accuracy~\cite{Amhis:2019ckw}. The grey solid (dashed) lines correspond to the estimated sensitivity of $4\%$ ($10\%$) precision on $\Gamma(B_c^+\to \tau^+\nu_\tau)/|V_{cb}|^2$ at FCC-ee.}
\label{fig:RD-RDst-gvl} 
\end{figure}

Further motivation to study $B_c^+\to\tau^+\nu_\tau$ decays at FCC-ee comes from the discrepancies observed in measurements of semileptonic $B$-meson decays based on the $b\to c \ell \nu_\ell$ transitions at LHCb and the $B$-factories~\cite{Lees:2013uzd,Aaij:2015yra,Aaij:2017deq,Aaij:2017tyk,Huschle:2015rga,Hirose:2016wfn,Hirose:2017dxl,Belle:2019rba}. More specifically, there is a combined $\approx 3.1~\sigma$ deviation between the experimental average of the lepton flavor universality ratios,
\begin{equation}
R_{D^{(\ast)}} = \dfrac{\mathcal{B}(B\to D^{(\ast)}\tau \bar{\nu})}{\mathcal{B}(B\to D^{(\ast)} l\bar{\nu})}\Bigg{\vert}_{l=e,\mu}\,,   
\end{equation}
with respect to their SM predictions, as discussed in~\cite{Amhis:2019ckw} and references therein. Similar discrepancies have also been observed in the ratio $R_{J/\psi}= \mathcal{B}(B_c^+ \to J/\psi \tau^+ \nu_\tau)/\mathcal{B}(B_c^+\to J/\psi \mu^+ \nu_\mu)$~\cite{Aaij:2017tyk}, but with lower statistical significance and still limited experimental precision. These discrepancies can be simultaneously explained by NP contributions to the effective coefficients defined in Eq.~\eqref{eq:left}. The simplest of these explanations requires an operator with the same chirality as the SM operator, with an effective coupling within the following $1\sigma$ range,
\begin{equation}
\label{eq:gVL}
g_{V_L} \in (0.05,0.09)\,,    
\end{equation}
as detailed in~\cite{Angelescu:2021lln} and references therein. Such an effective scenario could be induced by the tree-level exchange of the vector leptoquark $U_1=(\mathbf{3},\mathbf{1},2/3)$ or the scalar $S_1=(\mathbf{\bar{3}},\mathbf{1},1/3)$, with couplings to left-handed fermions~\cite{Angelescu:2021lln}. These particles are written in terms of their SM quantum numbers, $(SU(3)_c,SU(2)_L,U(1)_Y)$, with the convention $Q=Y+T_3$ for the electric charge, where $Y$ denotes the hypercharge and $T_3$ the third component of weak isospin. The effective couplings in Eq.~\eqref{eq:gVL} would imply a deviation from the SM prediction in $\Gamma(B_c^+\to \tau^+ \nu_\tau)/|V_{cb}|^2$ larger than $\mathcal{O}(10\%)$, which could be fully probed at FCC-e, as depicted in Fig.~\ref{fig:RD-RDst-gvl}. 

\begin{figure}[!t]
\centering
\includegraphics[width=.5\linewidth]{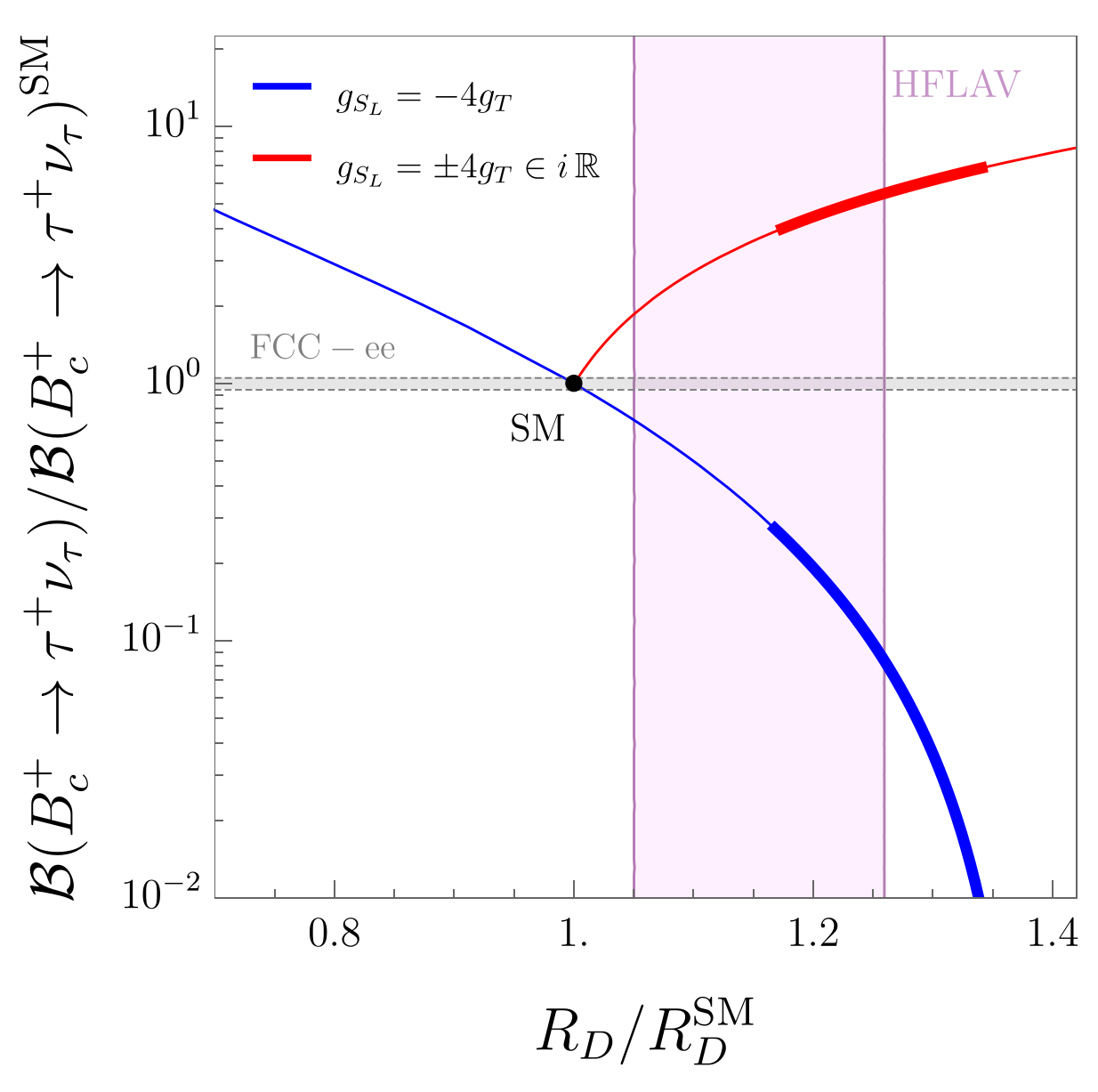}\put(-160,50){(a)}
\includegraphics[width=.5\linewidth]{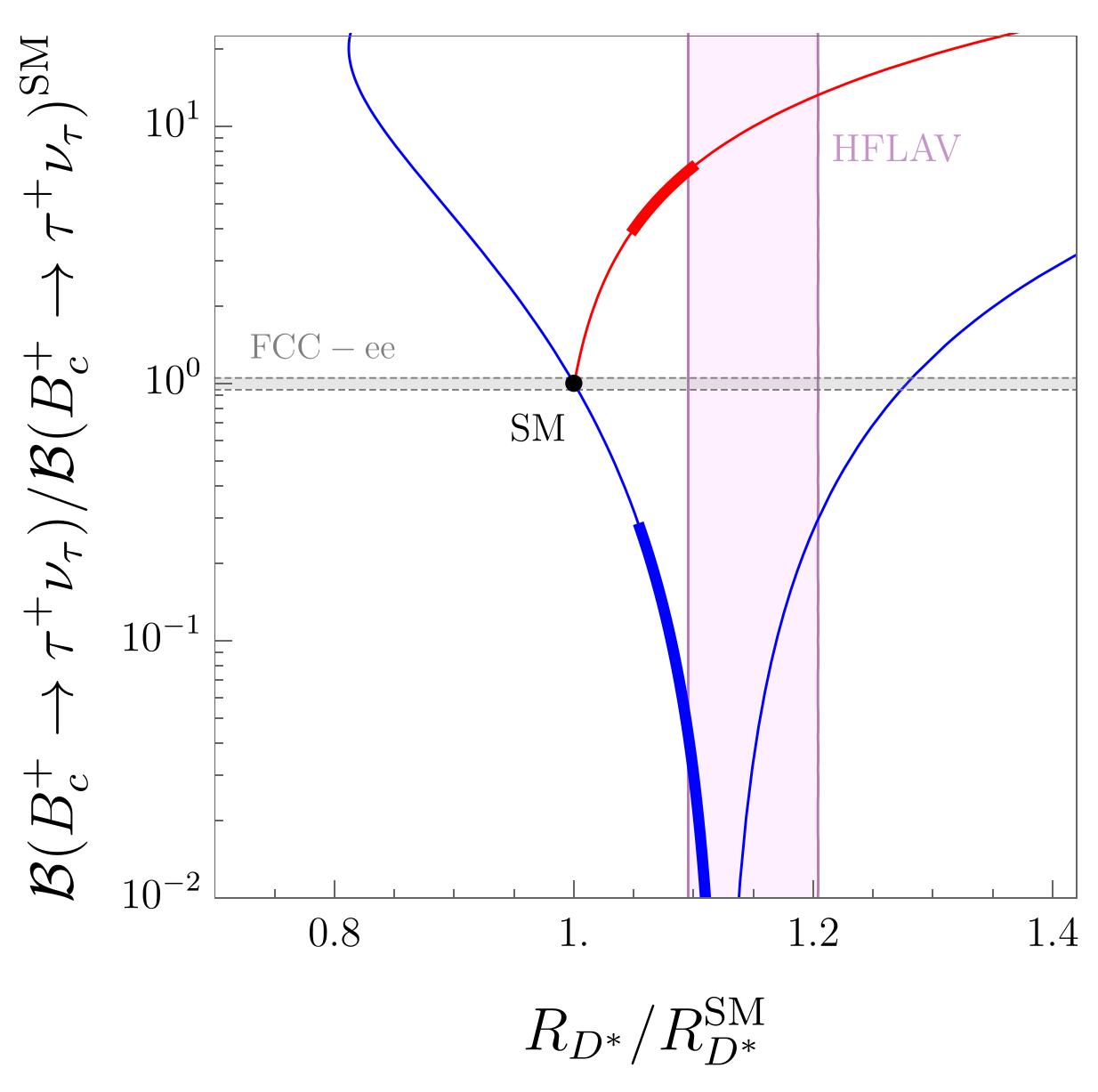}\put(-160,50){(b)}
\caption{Predictions for (a) $R_D/R_D^\mathrm{SM}$ and (b) $R_{D^\ast}/R_{D^\ast}^\mathrm{SM}$ are plotted against the ratio \mbox{$\mathcal{B}(B_c^+\to\tau^+ \nu_\tau)/\mathcal{B}(B_c^+\to\tau^+ \nu_\tau)^\mathrm{SM}$} in the effective scenarios: (i) $g_{S_L}=- 4 g_T$ (blue) and (ii) $g_{S_L}=+ 4 g_T\in i\,\mathbb{R}$ (red), which are defined at $\Lambda \approx 1~\mathrm{TeV}$. The thick lines correspond to the values of the effective couplings favoured by the current fit to $b\to c\tau \nu_\tau$ data~\cite{Angelescu:2021lln}. The magenta shaded regions denote the current experimental averages of $R_D$ and $R_{D^\ast}$ at $1\sigma$ accuracy~\cite{Amhis:2019ckw}. The grey region corresponds to the estimated sensitivity of $10\%$ precision on $\Gamma(B_c^+\to \tau^+\nu_\tau)/|V_{cb}|^2$ at FCC-ee.}
\label{fig:RD-RDst} 
\end{figure}

Another viable possibility to explain the discrepancies in $R_{D^{(\ast)}}$ are specific combinations of scalar and tensor operators, which are predicted in certain leptoquark models~\cite{Sakaki:2013bfa,Feruglio:2018fxo,Becirevic:2018afm,Becirevic:2018uab,Angelescu:2018tyl,Crivellin:2019dwb,Gherardi:2020qhc,Saad:2020ucl}. More precisely, the tree-level matching of the scalar leptoquarks with quantum numbers $R_2=(\mathbf{3},\mathbf{2},7/6)$ and $S_1=(\mathbf{\bar{3}},\mathbf{1},1/3)$ to the effective Hamiltonian \eqref{eq:left} can induce the combinations of couplings $g_{S_L}(\Lambda)= +4 g_T(\Lambda)$ and $g_{S_L}(\Lambda)= -4 g_T(\Lambda)$, respectively, at the matching scale $\Lambda \approx 1~\mathrm{TeV}$. After accounting for the renormalisation group running from $\approx 1~\mathrm{TeV}$ to $m_b$~\cite{Gonzalez-Alonso:2017iyc}, these relations become $g_{S_L}(m_b)\approx +8.1 g_T(m_b)$ and $g_{S_L}(m_b)\approx -8.5 g_T(m_b)$, respectively, which are known to provide a good description of $b\to c\tau \nu_\tau$ data~\cite{Angelescu:2021lln}. The latter scenario can explain the discrepancies in $R_D$ and $R_{D^\ast}$ via purely real effective couplings, whereas the first requires purely imaginary values. The allowed ranges for these couplings are given for example in Ref.~\cite{Angelescu:2021lln}. The impact of a measurement of $B_c^+\to \tau^+ \nu_\tau$ decays is even more dramatic for these scenarios due to the presence of a nonzero $g_{S_L}$ coupling, which induces sizeable contributions to $\mathcal{B}(B_c\to \tau^+ \nu_\tau)$ for the couplings required to explain $R_{D^{(\ast)}}$, as shown in Fig.~\ref{fig:RD-RDst}.

Clearly, a determination of $\Gamma (B_c^+\to \tau^+ \nu_\tau)/|V_{cb}|^2$ with $4\%$ precision can fully probe all of the leptoquark explanations of $R_D$ and $R_{D^\ast}$, as shown by the thick lines in Figs.~\ref{fig:RD-RDst-gvl} and \ref{fig:RD-RDst}. Since leptoquarks are the only viable explanations of these discrepancies which are still consistent with various flavor observables~\cite{Angelescu:2021lln,Buttazzo:2017ixm,Cornella:2021sby}, high-$p_T$ LHC limits~\cite{Faroughy:2016osc,Greljo:2018tzh,Hiller:2018wbv,Schmaltz:2018nls,Alves:2018krf,Angelescu:2021lln}, and electroweak precision constraints~\cite{Feruglio:2016gvd,Feruglio:2017rjo,Cornella:2018tfd}, the analysis presented within demonstrates the unique potential of a $B_c^+ \to \tau^+ \nu_\tau$ measurement at the FCC-ee $Z$-pole to either confirm or refute these anomalies.

\section{Conclusion}
\label{sec:conclusion}
In this work, the prospects for a precise measurement of the branching fraction of \BcToTauNu decays at the FCC-ee $Z$-pole are evaluated. Large simulated samples of \BcToTauNu signal and associated backgrounds are generated using the FCC-ee software to simulate detector effects, reconstruct signal candidates, and perform selection optimisation and fit studies. A two-stage BDT selection is employed to reduce all sources of hadronic $Z$ background, first using topological event-level information, and then the vertex properties of the detached $\TauThreePi$ decay to reduce the rate of $b$-hadron backgrounds. 
The sensitivities for both the branching fraction of \BcToTauNu and the ratio $R_c=\mathcal{B}(B_c^+\to \tau^+ \nu_\tau)/\mathcal{B}(B_c^+\to J/\psi \mu^+ \nu_\mu)$ are estimated as a function of the number of collected $Z$ decays, where a relative precision of around 4\% is achieved for $R_c$ with $N_Z = 5 \times 10^{12}$. The precision on the absolute branching fraction is limited to around 8\% due to knowledge of the $B_c^+ \to J/\psi \mu^+ \nu_\mu$ decay form factors, which can be improved through dedicated measurements of this mode in future.

The impact of a measurement of $B_c^+\to\tau^+ \nu_\tau$ on NP scenarios is also discussed. In particular, it is shown that such a measurement at FCC-ee can constrain a large region of the $(\tan \beta, m_{H^\pm})$ plane in the type-II 2HDM, which cannot be covered by other flavour-physics measurements.  Recently, leptoquark models have received significant attention as the only viable explanation of the $B$-physics anomalies in both charged and neutral current processes. A precise measurement of the branching fraction of \BcToTauNu at FCC-ee could fully probe the interpretations of $R_D$ and $R_{D^\ast}$ that are permitted under existing constraints.

In summary, this work demonstrates why FCC-ee is the most well-suited environment for a measurement of the branching fraction of the \BcToTauNu decay, and represents the first FCC-ee analysis to use common software tools from \edmhep through to final analysis.

\section*{Acknowledgements}
The authors would like to thank  D.~Bečirević, M.~John, S.~Monteil, P.~Robbe, and M.-H.~Schune for the useful discussions and their input. 
This project has received support from the European Union’s Horizon 2020 research and innovation programme under the Marie Skłodowska-Curie grant agreement N$^\circ$~860881-HIDDeN.

\clearpage
\appendix

\section{Exclusive background efficiencies}
\label{app:bkg_effs}

\renewcommand{\arraystretch}{1.5}{
\begin{table}[h!]
\centering
\scriptsize
\begin{tabular}{lllll}
Decay mode & N(expected) & N(generated) & Expected / Generated & Final $\epsilon$ \\ \hline
$B^+ \to \bar{D}^0 \tau^+ \nu_\tau$ & $5.01 \times 10^{9}$ & $2 \times 10^{8}$ & $25.0$ & $1.46 \times 10^{-9}$ \\
$B^+ \to \bar{D}^{*0} \tau^+ \nu_\tau$ & $1.22 \times 10^{10}$ & $2 \times 10^{8}$ & $61.1$ & $1.1 \times 10^{-9}$ \\
$B^+ \to \bar{D}^0 3\pi$ & $3.64 \times 10^{9}$ & $1.9 \times 10^{8}$ & $19.2$ & $1.56 \times 10^{-9}$ \\
$B^+ \to \bar{D}^{*0} 3\pi$ & $6.7 \times 10^{9}$ & $2 \times 10^{8}$ & $33.5$ & $1.04 \times 10^{-9}$ \\
$B^+ \to \bar{D}^0 D_s^+$ & $5.85 \times 10^{9}$ & $2 \times 10^{8}$ & $29.3$ & $2.52 \times 10^{-10}$ \\
$B^+ \to \bar{D}^{*0} D_s^+$ & $4.94 \times 10^{9}$ & $1.75 \times 10^{8}$ & $28.2$ & $2.72 \times 10^{-10}$ \\
$B^+ \to \bar{D}^{*0} D_s^{*+}$ & $1.11 \times 10^{10}$ & $2 \times 10^{8}$ & $55.6$ & $2.42 \times 10^{-10}$ \\
\hline
$B^0 \to D^- \tau^+ \nu_\tau$ & $7.02 \times 10^{9}$ & $2 \times 10^{8}$ & $35.1$ & $2.69 \times 10^{-9}$ \\
$B^0 \to D^{*-} \tau^+ \nu_\tau$ & $1.02 \times 10^{10}$ & $2 \times 10^{8}$ & $51.0$ & $1.25 \times 10^{-9}$ \\
$B^0 \to D^- 3\pi$ & $3.9 \times 10^{9}$ & $2 \times 10^{8}$ & $19.5$ & $3.4 \times 10^{-9}$ \\
$B^0 \to D^{*-} 3\pi$ & $4.69 \times 10^{9}$ & $2 \times 10^{8}$ & $23.4$ & $9.84 \times 10^{-10}$ \\
$B^0 \to D^- D_s^+$ & $4.68 \times 10^{9}$ & $2 \times 10^{8}$ & $23.4$ & $3.23 \times 10^{-10}$ \\
$B^0 \to D^{*-} D_s^+$ & $5.2 \times 10^{9}$ & $2 \times 10^{8}$ & $26.0$ & $2.32 \times 10^{-10}$ \\
$B^0 \to D^{*-} D_s^{*+}$ & $1.15 \times 10^{10}$ & $2 \times 10^{8}$ & $57.5$ & $2.35 \times 10^{-10}$ \\
\hline
$B_s^0 \to D_s^- \tau^+ \nu_\tau$ & $3.53 \times 10^{9}$ & $2 \times 10^{8}$ & $17.6$ & $3.71 \times 10^{-9}$ \\
$B_s^0 \to D_s^{*-} \tau^+ \nu_\tau$ & $2.35 \times 10^{9}$ & $2 \times 10^{8}$ & $11.8$ & $2.27 \times 10^{-9}$ \\
$B_s^0 \to D_s^- 3\pi$ & $8.85 \times 10^{8}$ & $2 \times 10^{8}$ & $4.4$ & $5.53 \times 10^{-9}$ \\
$B_s^0 \to D_s^{*-} 3\pi$ & $1.05 \times 10^{9}$ & $2 \times 10^{8}$ & $5.2$ & $3.38 \times 10^{-9}$ \\
$B_s^0 \to D_s^- D_s^+$ & $6.39 \times 10^{8}$ & $2 \times 10^{8}$ & $3.2$ & $4.09 \times 10^{-10}$ \\
$B_s^0 \to D_s^{*-} D_s^+$ & $2.02 \times 10^{9}$ & $2 \times 10^{8}$ & $10.1$ & $3.17 \times 10^{-10}$ \\
$B_s^0 \to D_s^{*-} D_s^{*+}$ & $2.09 \times 10^{9}$ & $2 \times 10^{8}$ & $10.5$ & $2.56 \times 10^{-10}$ \\
\hline
$\Lambda_b^0 \to \Lambda_c^- \tau^+ \nu_\tau$ & $1.83 \times 10^{9}$ & $2 \times 10^{8}$ & $9.1$ & $1.36 \times 10^{-9}$ \\
$\Lambda_b^0 \to \Lambda_c^{*-} \tau^+ \nu_\tau$ & $1.83 \times 10^{9}$ & $2 \times 10^{8}$ & $9.1$ & $9.44 \times 10^{-10}$ \\
$\Lambda_b^0 \to \Lambda_c^- 3\pi$ & $4.31 \times 10^{8}$ & $2 \times 10^{8}$ & $2.2$ & $5.58 \times 10^{-9}$ \\
$\Lambda_b^0 \to \Lambda_c^{*-} 3\pi$ & $4.31 \times 10^{8}$ & $2 \times 10^{8}$ & $2.2$ & $9.21 \times 10^{-10}$ \\
$\Lambda_b^0 \to \Lambda_c^- D_s^+$ & $6.15 \times 10^{8}$ & $2 \times 10^{8}$ & $3.1$ & $3.46 \times 10^{-10}$ \\
$\Lambda_b^0 \to \Lambda_c^{*-} D_s^+$ & $6.15 \times 10^{8}$ & $2 \times 10^{8}$ & $3.1$ & $2.72 \times 10^{-10}$ \\
$\Lambda_b^0 \to \Lambda_c^{*-} D_s^{*+}$ & $6.15 \times 10^{8}$ & $2 \times 10^{8}$ & $3.1$ & $2.5 \times 10^{-10}$ \\
\hline
\end{tabular}
\caption{Summary of the exclusive $B$-hadron background samples used for determination of the optimal BDT cuts and to model background in the signal yield fit. The yields expected for each decay mode with $N_Z = 5 \times 10^{12}$ are shown in the second column, and the generated sample statistics are shown in the third column. The large ratio between expected and generated statistics (fourth column) illustrates why cut-and-count efficiencies cannot be used to determine the expected background rejection achieved by the double-BDT selection; for many background modes, none of the generated events survive the optimal BDT cuts applied. The efficiency values determined using the spline parameterisation approach are given in the final column. Using splines derived from the total exclusive sample enables efficiencies down to the $10^{-10}$ level to be evaluated.}
\label{tab:bkg_BDT_effs}
\end{table}
}

\FloatBarrier

\newpage

\section{Signal yield and branching fraction precision estimates}
\label{app:precisions}

\renewcommand{\arraystretch}{1.4}{
\begin{table}[h!]
\centering
\small
\begin{tabular}{ll}
$N_Z (\times 10^{12})$ & Relative $\sigma$ ($\sigma_{syst}^N = [0, 0.25, 0.5, 1] \times \sigma_{stat}^N$) \\ \hline
0.5 & [0.078, 0.087, 0.096, 0.111] \\
1 & [0.055, 0.061, 0.067, 0.077] \\
2 & [0.038, 0.042, 0.046, 0.053] \\
3 & [0.032, 0.036, 0.039, 0.046] \\
4 & [0.027, 0.03, 0.033, 0.038] \\
5 & [0.024, 0.027, 0.03, 0.034] \\
\hline
\end{tabular}
\caption{Estimated relative precision on $N(B_c^+ \to \tau^+ \nu_\tau)$ as a function of $N_Z$, where four different levels of systematic uncertainty on the signal yield are shown.}
\label{tab:N_Bc2TauNu_vs_NZ}
\end{table}
}

\renewcommand{\arraystretch}{1.4}{
\begin{table}[h!]
\centering
\small
\begin{tabular}{ll}
$N_Z (\times 10^{12})$ & Relative $\sigma$ ($\sigma_{syst}^N = [0, 0.25, 0.5, 1] \times \sigma_{stat}^N$) \\ \hline
0.5 & [0.081, 0.09, 0.098, 0.113] \\
1 & [0.058, 0.064, 0.07, 0.08] \\
2 & [0.042, 0.046, 0.05, 0.056] \\
3 & [0.037, 0.04, 0.043, 0.049] \\
4 & [0.032, 0.035, 0.037, 0.042] \\
5 & [0.03, 0.032, 0.034, 0.038] \\
\hline
\end{tabular}
\caption{Estimated relative precision on $R_c$ as a function of $N_Z$, where four different levels of systematic uncertainty on the signal yield are shown.}
\label{tab:BF_ratio_vs_NZ}
\end{table}
}

\renewcommand{\arraystretch}{1.4}{
\begin{table}[h!]
\centering
\small
\begin{tabular}{ll}
$N_Z (\times 10^{12})$ & Relative $\sigma$ ($\sigma_{syst}^N = [0, 0.25, 0.5, 1] \times \sigma_{stat}^N$) \\ \hline
0.5 & [0.115, 0.122, 0.128, 0.139] \\
1 & [0.1, 0.104, 0.107, 0.114] \\
2 & [0.092, 0.093, 0.095, 0.099] \\
3 & [0.089, 0.091, 0.092, 0.095] \\
4 & [0.088, 0.089, 0.09, 0.092] \\
5 & [0.087, 0.088, 0.088, 0.09] \\
\hline
\end{tabular}
\caption{Estimated relative precision on $\mathcal{B}(B_c^+ \to \tau^+ \nu_\tau)$ as a function of $N_Z$, where four different levels of systematic uncertainty on the signal yield are shown.}
\label{tab:BF_Bc2TauNu_vs_NZ}
\end{table}
}

\FloatBarrier

\bibliographystyle{LHCb}
\bibliography{references}

\end{document}